\documentclass[aps,pra,reprint,twocolumn,superscriptaddress,amsmath,amssymb,citeautoscript]{revtex4-1}
\usepackage{amsmath}
\usepackage{amssymb}
\usepackage{graphicx}
\usepackage{color}
\usepackage{bm}
\usepackage{epstopdf}
\usepackage{grffile}
\DeclareGraphicsExtensions{.eps}

\usepackage{tabularx}
\usepackage{color, colortbl}
\definecolor{Gray}{gray}{0.9}
\makeatletter
\def\CT@@do@color{%
\global\let\CT@do@color\relax
      \@tempdima\wd\z@
      \advance\@tempdima\@tempdimb
      \advance\@tempdima\@tempdimc
\advance\@tempdimb\tabcolsep
\advance\@tempdimc\tabcolsep
\advance\@tempdima2\tabcolsep
      \kern-\@tempdimb
      \leaders\vrule
              \hskip\@tempdima\@plus  1fill
      \kern-\@tempdimc
      \hskip-\wd\z@ \@plus -1fill }
\makeatother

\newcommand{\e}{\rm{e}}
\newcommand{\ii}{\rm{i}}

\begin{document}

\title{Symmetry-broken states in a system of interacting bosons on a two-leg ladder with a uniform Abelian gauge field}
\author{S. Greschner}
\affiliation{Institut f\"ur Theoretische Physik, Leibniz Universit\"at Hannover, 30167~Hannover, Germany} 
\author{M. Piraud} 
\affiliation{Department of Physics and Arnold Sommerfeld Center for Theoretical Physics, Ludwig-Maximilians-Universit\"at M\"unchen, 80333 M\"unchen, Germany}
\author{F. Heidrich-Meisner}
\affiliation{Department of Physics and Arnold Sommerfeld Center for Theoretical Physics, Ludwig-Maximilians-Universit\"at M\"unchen, 80333 M\"unchen, Germany}
\affiliation{Kavli Institute for Theoretical Physics, University of California, Santa Barbara CA 93106, USA}
\author{I. P. McCulloch}
\affiliation{ARC Centre for Engineered Quantum Systems, School of Mathematics and Physics, The University of Queensland, St Lucia, QLD 4072, Australia}
\author{U. Schollw\"ock}
\affiliation{Department of Physics and Arnold Sommerfeld Center for Theoretical Physics, Ludwig-Maximilians-Universit\"at M\"unchen, 80333 M\"unchen, Germany}
\author{T. Vekua}
\affiliation{Institut f\"ur Theoretische Physik, Leibniz Universit\"at Hannover, 30167~Hannover, Germany} 
\affiliation{James Franck Institute, The University of Chicago, Chicago IL 60637, USA}

\begin{abstract}
We study the quantum phases of bosons with repulsive contact interactions on a two-leg ladder in the presence of a uniform Abelian gauge field. The model realizes many interesting states, including Meissner phases, vortex-fluids, vortex-lattices, charge-density-waves and the biased-ladder phase. Our work focuses on the subset of these states that break a discrete symmetry. We use density matrix renormalization group simulations to demonstrate the existence of three vortex-lattice states at different vortex densities and we characterize the phase transitions from these phases into neighboring states. Furthermore, we provide an intuitive explanation of the chiral-current reversal effect that is tied to some of these vortex lattices. We also study a charge-density-wave state that exists at 1/4 particle filling at large interaction strengths and flux values close to half a flux quantum. By changing the system parameters, this state can transition into a completely gapped vortex-lattice Mott-insulating state. We elucidate the stability of these phases against nearest-neighbor interactions on the rungs of the ladder relevant for experimental realizations with a synthetic lattice dimension. A charge-density-wave state at 1/3 particle filling can be stabilized for flux values close to half a flux-quantum and for very strong on-site interactions in the presence of strong repulsion on the rungs. Finally, we analytically describe the emergence of these phases in the low-density regime, and, in particular, we obtain the boundaries of the biased-ladder phase, i.e., the phase that features a density imbalance between the legs. We make contact to recent quantum-gas experiments that realized related models and discuss signatures of these quantum states in experimentally accessible observables.
\end{abstract}
\date{\today}

\maketitle
\section{Introduction}

An important part of the physics of quantum particles moving in a two-dimensional plane under the action of a magnetic field is related to the quantum Hall effect~\cite{Klitzing1980,Tsui1982}. While the essence of the integer quantum Hall effect can be understood from considering  noninteracting electrons, interacting  particles in two dimensions and in the presence of an Abelian gauge field provide an ideal playground to explore exotic many-body physics~\cite{Laughlin1983,Moore1991,Cooper2013}, encompassing, most notably, the fractional quantum Hall effect.
 Periodic lattice potentials introduce additional intriguing physics. A magnetic field applied perpendicular to the plane of motion of a charged particle in a lattice produces a fascinating structure of energy levels known as the Hofstadter butterfly~\cite{Hofstadter1976} resolved experimentally in solid state systems only recently \cite{Hunt2013,Dean2013,Ponomarenko2013}.

The experimental progress in the field of ultracold atomic gases with emulating gauge fields or spin-orbit coupling 
in these systems of neutral particles has opened new prospects for observing many-body physics in the presence
of gauge fields in a very clean and highly tunable environment \cite{Dalibard2011,Galitski2013,Goldman2016}.
While pioneering experiments have demonstrated the successful implementation of spin-orbit coupling in Bose gases in the 
continuum \cite{Lin2009,Lin2011}, the field has seen a tremendous activity in studying the combined effects of optical lattices with artificially engineered gauge fields \cite{Jimenez2012,Struck2012,aidelsburger2011,Aidelsburger2013,Miyake2013}, accessing the physics of the Hofstadter model \cite{Aidelsburger2013,Miyake2013} as well as the famous Haldane model \cite{Haldane1988,Jotzu2014}.

So far, quantum gas experiments have focussed on non or weakly-interacting quantum gases. 
In this regime,  some key hallmark features of topological states of matter  \cite{Hasan2010,Qi2011} were measured, such as the Zak phase in one-dimensional systems \cite{Atala2013}, the Berry curvature in a Floquet system \cite{Flaeschner2016}, the Chern number \cite{Aidelsburger2015,Jotzu2014} or the Berry flux in momentum space \cite{Duca2015}.

The strongly-interacting regime in conjunction with artificial gauge fields, though, remains largely unexplored by ultracold atomic gases experiments. Theoretically, numerous studies have addressed the interplay of interactions, gauge fields and lattice topology and have made many predictions for exciting physics that could be observed with quantum gases in optical lattices. These include (interacting) Chern insulators of both fermions and bosons \cite{Moeller2009,Varney2010,Varney2011,Kjaell2012,He2015}, fractional Chern insulators \cite{Sorensen2005,Regnault2011,Neupert2011,Moeller2015}, and exotic forms of quantum magnetism \cite{Lim2008, Orth2012,Radic2012,Cole2012,Zhou2013,Piraud2014}, to name but a few examples (see recent reviews for a more comprehensive overview \cite{Galitski2013,Goldman2016}).

\begin{figure}[b]
\includegraphics[width=1\linewidth]{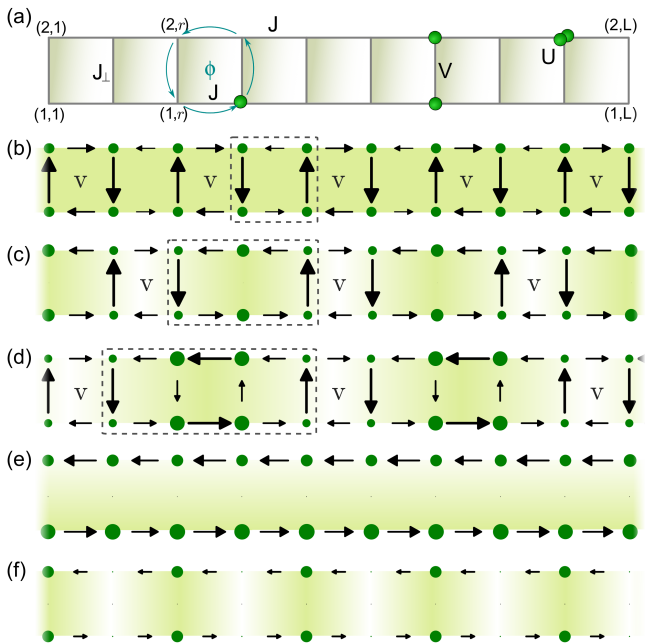}
\caption{(a) Sketch of the two-leg ladder model and interaction- and tunneling terms as defined in Eqs.~\eqref{eq:ham_rung}, ~\eqref{eq:ham_leg} and ~\eqref{eq:ham_Vrung}. Current patterns and onsite density in (b)-(d) the different vortex lattices with vortex density (b)   $\rho_v=1/2$, (c)   $\rho_v=1/3$, and (d)   
$\rho_v=1/4$, (e) the biased-ladder phase (BLP) phase, and (f) in the charge-density-wave (CDW$_{1/4}$) phase at particle density $\rho=1/4$. The arrows indicate the direction and by their length, the strength
of local currents. The density is represented by the size of the circles and the background shading. 
}
\label{fig:patterns}
\end{figure}

Our work is primarily motivated by the experimental realization of {\it ladder} systems combined with uniform Abelian gauge fields in ultracold atomic gases.
Ladders, which here we have in mind to be arrays of plaquettes as indicated in Fig.~\ref{fig:patterns}, are the simplest possible
extensive lattices that allow one to study nontrivial orbital effects in the presence of a synthetic magnetic field.  
Such ladder structures can either be obtained using superlattices or a so-called synthetic lattice dimension \cite{Celi2014}.
The former approach has been utilized in \cite{Atala2014} to study a weakly-interacting Bose gas loaded into an optical lattice with the two-leg ladder geometry. This experiment heavily relied on the measurement of local currents and could  access the chiral edge current 
predicted to exist in this system in the presence of a uniform flux per unit cell \cite{Orignac2001,Huegel2014}. Moreover, the experiment
established a similarity to superconductors (also previously discussed in theoretical papers \cite{Orignac2001,Huegel2014}),
since the chiral current at small flux behaves similar to the screening current in the Meissner phase (with no current in the bulk, which, for a ladder, implies a vanishing of 
local currents on its rungs) while at large fluxes, finite rung currents emerge, reminiscent of a vortex phase in type-II superconductors.

The synthetic lattice dimension approach combines an actual optical lattice, which is typically one-dimensional, with Raman lasers that drive
transitions between a subset of the hyperfine states of bosonic or fermionic atoms \cite{Celi2014}.  In the first two implementations of this scheme, two- and three-leg ladders have been realized using this method \cite{Mancini2015,Stuhl2015}, and both experiments also succeeded in measuring edge currents (more specifically momentum distributions), for fermions \cite{Mancini2015} and bosons \cite{Stuhl2015}.
More recent experiments realized two-leg ladders using an optical clock transition \cite{Livi2016,Kolkowitz2016} (based on the proposal from \cite{Wall2016}) or even using an  all-synthetic-lattice approach \cite{Alex2016}.

These examples demonstrate the experimental study of the effects of synthetic gauge fields in neutral ultracold atomic gases in low-dimensional lattice geometries constitutes an active field of timely research. Despite the enormous recent progress with experimental observations of fascinating single-particle behavior due to synthetic gauge fields, the strongly-interacting regime remains hard to access. One possible reason
that is being investigated is  heating in driven many-body systems  \cite{Dalessio2012,Dalessio2014,Lazarides2014,Bukov2016} due to the various driving schemes used to mimic the artificial gauge fields \cite{Aidelsburger2013,Miyake2013,Struck2012,Jotzu2014}. 
 
The  interest in bosonic ladders dates back to the development of the theory of arrays of Josephson junctions \cite{Kardar1986,Granato1990,Mazo1995,Denniston1995}, adequately described by high density and weak interactions. Subsequently, a first bosonization study for weakly-coupled legs of a bosonic two-leg ladder
explored the strongly-interacting regime, predicting the stability of Meissner-like, vortex fluids and vortex lattices \cite{Orignac2001},
whose existence was  initially established for the weakly-interacting regime only \cite{Kardar1986,Granato1990,Mazo1995,Denniston1995}.
Then, fueled by the experimental progress with emulating artificial gauge fields, Dhar {\it et al.}~\cite{Dhar2012,Dhar2013} demonstrated the existence of so-called chiral Mott insulators in bosonic two-leg ladders at flux $\phi=\pi$ per plaquette (see the sketch of the model in Fig.~\ref{fig:patterns}(a)), which spontaneously break time-reversal symmetry. 
A bosonization study of the orbital response of a bosonic ladder in the strongly-interacting regime at arbitrary flux was presented in \cite{Petrescu2013},
predicting Meissner and vortex phases also in the Mott-insulating case.
Understanding the orbital response of interacting fermions on two-leg ladders has been the topic of Refs.~\cite{Roux2007,Carr2006}.

The experiments on bosonic \cite{Atala2014,Stuhl2015} and fermionic ladders \cite{Mancini2015}
have led to numerous theoretical studies of the strongly-correlated quantum phases of such systems \cite{Keles2015,Piraud2015,Grusdt2014,Cornfeld2015,Petrescu2015,DiDio2015,DiDio2015a,Barbarino2015,Barbarino2016,Wei2014,Uchino2015}.
A particular interest has been in the possible existence of one-dimensional versions of fractional quantum Hall states \cite{Grusdt2014,Cornfeld2015,Petrescu2015,Barbarino2015}.
Whether  topologically nontrivial states that resemble properties of, e.g., the $\nu=1/2$ Laughlin state are indeed realized in bosonic or fermionic ladders remains a topic of ongoing research \cite{Petrescu-aps}.

Other efforts addressed the fate of the Meissner and vortex-like states inherited from the noninteracting limit of bosonic two-leg ladders.
Their origin can be traced back to the existence of either a unique minimum in the single-particle dispersion at quasi-momentum  or 
two equivalent minima at incommensurate momenta, the former corresponding to the Meissner, the latter to the vortex phase \cite{Huegel2014,Atala2014}.
Similar states survive even in the limit of strongly-interacting bosons \cite{Petrescu2013,Keles2015,Piraud2015}, both on top of Mott-insulators and superfluids. Overall, interactions lead to a suppression of vortex phases.

While most of the theoretical work has investigated the case of pure contact interactions of the type $H_{\rm int} = U/2 \sum_i n_i (n_i-1)$, where $n_i$ measures the density in 
a site of a lattice ($U$ is the interaction strength), in the experimental realizations using a synthetic lattice dimension \cite{Stuhl2015,Mancini2015}, atoms experience 
long-range interactions. The reason is that the physical interaction is of contact type, meaning that all atoms in a {\it rung} interact with each other, regardless 
of their hyperfine state. The effects of longer-range interactions have been studied in \cite{Kolley2015,Zeng2015,yan15,ghosh15,Barbarino2015,Natu2015,Petrescu2015,Bilitewski2016,taddia2016,Ghosh2016,Anisimovas2016}.

Returning now to the case of bosonic two-leg ladders with contact interactions in the model, while Meissner-like and vortex-liquid phases
exist in the presence of strong interactions, 
vortex lattices, on the contrary, 
have remained elusive in the strongly-interacting regime, with the exception of the chiral Mott insulators \cite{Dhar2012,Dhar2013}, which in fact represents a vortex lattice state with a maximal possible vortex density  $\rho_v=1/2$, and a unit cell of two plaquettes as shown in \cite{Greschner2015}. In Ref.~\cite{Greschner2015}, we determined the range of stability of the above mentioned vortex lattice at vortex densities of $\rho_v=1/2$ and of another vortex lattice state $\rho_v=1/3$ (having a unit cell made of three plaquettes) in the regime of intermediate interactions $1< U/J<10$ and a low  filling $\rho \leq 1 $.
Moreover, we discovered that interactions can lead to a spontaneous reversal of the circulation direction of the boundary chiral current in certain 
vortex-lattice states: the atoms there behave as if the direction of the external magnetic field (or the sign of the flux) had been inverted. 
This effect can be understood as resulting from the periodicity of the chiral current with $2\pi$, the increase of the effective flux from $\phi$ to $q \phi$ in vortex lattices with unit cells of $q$ plaquettes and the quantum nature of atoms. Only those vortex lattices that are stable at values of the flux for which $\pi <q \phi < 2 \pi$ lead to this reversal. Remarkably, this reversal is stable against temperatures that are possible to realize in current experiments and hence it can be used as an experimental probe of the existence
of vortex lattices. In bosonic ladders, such a reversal of the chiral current occurs only in the interacting case, while for fermions, in a complicated band structure that results from adding the flux, changing the filling can lead to a chiral-current reversal already in noninteracting systems simply due to the Pauli principle \cite{Roux2007}.

The purpose of the present work is to provide a comprehensive analysis of quantum phases in bosonic two-leg ladders subject to a uniform flux $\phi$ per plaquette that spontaneously break some {\it discrete} symmetry of the microscopic model. We primarily focus on the low-density regime $\rho \leq 1$ and study such states as a function of interaction strength, flux, and 
the ratio $J_\perp/J$ of hopping matrix elements along rungs $J_\perp$ and legs $J$ (see the sketch of the model shown in Fig.~\ref{fig:patterns}(a)).

Most notably, the list of states with broken discrete symmetries includes vortex lattices, which break lattice translation invariance. We study the previously known vortex lattices at vortex densities
$\rho_v=1/2$ \cite{Dhar2012,Dhar2013,Greschner2015} and $1/3$ \cite{Greschner2015} and we report numerical evidence for the existence of an additional vortex lattice at $\rho_v=1/4$.
Typical configurations for the local particle currents and densities in these three vortex lattice states are plotted in Figs.~\ref{fig:patterns}(b)-(d). 
The position of the vortex cores is denoted by the symbol V in the figure. While the screening current in a Meissner phase goes counterclockwise around the boundary of the system, the vortices carry currents of the opposite chirality,
thus reducing the overall chiral current. In addition, the vortex lattices with $\rho_v<1/2$  feature density modulations that are locked to the structure of the vortex lattice.
The results shown in Fig.~\ref{fig:patterns}(b)-(f)  were obtained from density matrix renormalization group (DMRG) simulations \cite{White1992,Schollwoeck2011,Schollwoeck2005}, the primary tool in our analysis.

Our main results beyond those of Ref.~\cite{Greschner2015} for the vortex lattices are the discovery of a stable $\rho_v=1/4$ vortex lattice, the stability analysis of the $\rho_v=1/3$ against increasing interactions, 
the analysis of the phase transitions between the vortex lattices and the neighboring phases, and an intuitive discussion of the chiral-current reversal that we develop
by comparing systems with spontaneously enlarged unit cells to systems with explicitly larger unit cells.
Moreover, we demonstrate that nearest-neighbor interactions on the rungs of a two-leg ladder suppress vortex phases in favor of the Meissner state.
From a conceptual point of view, it is very important to compare different measures of the vortex density, which does not have a microscopic definition.
We also discuss the signatures of vortex lattices in the experimentally accessible quasi-momentum distribution function, comparing different gauges. 

Another state that breaks a discrete symmetry is the biased-ladder phase (BLP), first discussed by Wei and Mueller \cite{Wei2014} and also studied in 
\cite{Uchino2015,Uchino2016,Greschner2015}. In this state, the density between the two legs of the ladder is imbalanced, which serves as an order parameter.
Thus, this state breaks the $Z_2$ symmetry associated with inversion of the two legs together with inversion of the sign of flux.
A typical configuration of the local currents and density in the BLP state is shown in Fig.~\ref{fig:patterns}(e).
We determined the phase boundaries of the BLP phase at intermediate interaction strength  in \cite{Greschner2015} from accurate DMRG simulations, 
thus providing robust evidence for its existence beyond mean-field \cite{Wei2014} and bosonization \cite{Uchino2015,Uchino2016} predictions.
In our present work we also introduce a theory of the emergence of the Meissner state, vortex fluids and the biased-ladder phase based on the limit of a dilute Bose gas that some of us originally developed for the description of frustrated spin chains just below their saturation magnetization \cite{Arlego2011,Kolezhuk2012}. 

Finally, we discuss charge-density-wave (CDW$_\rho$) states. CDW$_{1/4}$ states exist, in particular, in the limit of hard-core bosons at a filling of $\rho=1/4$ and for sufficiently
large values of the hopping on the rungs $J_{\perp}>J$ \cite{Piraud2015,DiDio2015}. In Ref.~\cite{Piraud2015}, we provided a theoretical  
explanation for their existence based on a mapping to an effective spin-1/2 Hamiltonian valid in the regime of $J_\perp >J$. In the present work,
we primarily focus on the stability of the CDW$_{1/4}$ state against going to lower values of $U/J<\infty$. As a result, we find that the CDW$_{1/4}$ state survives down to $U/J \gtrsim 30$
at quarter filling $\rho=1/4$. At smaller values of $U/J$, the system transitions from the CDW$_{1/4}$ state into a Mott insulator that carries a vortex lattice with $\rho_v=1/2$.
The existence of this Mott insulator is very interesting since it results from the combined effects of interactions, flux and filling, 
unlike other Mott insulators that exist in the bosonic two-leg ladder \cite{Dhar2012,Petrescu2013,Keles2015,Piraud2015,Vekua2003} that can be traced back to the limit of $\phi=0$.
A typical configuration of the local currents and density in the CDW$_{1/4}$ state is shown in Fig.~\ref{fig:patterns}(f).
For strong interactions in the synthetic dimension $U/J\to\infty$ and $V/J\to\infty$ we observe the stability of a CDW$_{1/3}$ state, i.e.,  at filling $\rho =1/3$.
 
\begin{table}[tb]
\begin{tabular}{ccccccc}
\hline
\hline
	& \hspace{1.8cm} & $c$ & \hspace{0.2cm}$\rho_v$\hspace{0.2cm} & $q$ & \hspace{0.2cm}${\rm avg} |j_R|$\hspace{0.2cm} & $\Delta n$\\[0.1cm]
\hline
Meissner phase & M-SF & $1$ & $0$ & $1$ & $0$ & $0$\\
 & M-MI & $0$ & $0$ & $1$  & $0$ & $0$\\
vortex liquid & V-SF & $2$ & $>0$ & $1$ & $0$ & $0$\\
& V-MI & $1$ & $>0$ & $1$  & $0$ & $0$\\
\rowcolor{Gray}
vortex lattice & VL$_{1/2}$-SF & $1$ & $1/2$ & $2$   & $>0$ & $0$\\
\rowcolor{Gray}
&VL$_{1/3}$-SF & $1$ & $1/3$ & $3$   & $>0$ & $0$\\
\rowcolor{Gray}
&VL$_{1/4}$-SF & $1$ & $1/4$ & $4$   & $>0$ & $0$\\
\rowcolor{Gray}
&VL$_{1/2}$-MI & $0$ & $1/2$ & $2$   & $>0$ & $0$\\
\rowcolor{Gray}
&\dots&&&&&\\
\rowcolor{Gray}
charge-density-wave & CDW$_{1/3}$ & $0$ & $0$ & $3$  & $0$ & $0$\\
\rowcolor{Gray}
& CDW$_{1/4}$ & $0$ & $0$ & $2$  & $0$ & $0$\\
\rowcolor{Gray}
&\dots&&&&&\\
\rowcolor{Gray}
biased-ladder phase & BLP-SF & $1$ & 0 & $1$  & $0$ & $>0$\\
\hline
\hline
\end{tabular}
\caption{\label{tab:phases} Quantum phases of bosons with repulsive contact interactions on a two-leg ladder with a uniform Abelian gauge field studied  in this work. 
We list those states that we actually detected in our numerical simulations with no claim of exclusiveness as additional states exist,
which we do not explicitly discuss in this work.
Meissner, vortex liquid and vortex lattice phases exist either atop superfluid (SF) or Mott-insulating (MI) states. We also list  characteristic properties (see the 
text for details) such as the central charge $c$, counting the number of gapless modes, the vortex density $\rho_v$, the size of the effective unit cell of the groundstate $q$ (plaquettes), the average local rung current in the thermodynamic limit ${\rm avg} |j_R|$ (see Eq.~\eqref{eq:avgjr}) and the leg-density imbalance $\Delta n$ (see Eq.~\eqref{eq:blpdeltan}).
The ``shaded'' states break a discrete symmetry and are at the main focus of this study.
}
\end{table}

Table~\ref{tab:phases} provides an overview over the quantum phases 
that are realized in the bosonic two-leg ladder with onsite interactions and some of their characteristic properties. 
These include the Meissner superfluid (M-SF), the Meissner-Mott insulator (M-MI), vortex liquid superfluids (V-SF), and vortex liquids on top of Mott insulators (V-MI), vortex-lattice superfluids at a vortex density $\rho_v$ (V$_{\rho_v}$-SF) as well as the VL$_{1/2}$-MI that sits on top of a Mott-insulating state, the charge-density-wave state (CDW$_{1/4}$)
and the biased-ladder superfluid phase (BLP). Other states that have also been proposed to exist in this model \cite{Petrescu-aps,Petrescu2015} are not at the main focus of the work and are thus not included in the table.
The terminology for these phases as well as the acronyms already suggest the existence of transitions in two sectors: the charge sector, in which the Mott-insulator-to-superfluid transition takes place and the antisymmetric sector, in which the Meissner-to-vortex-to-vortex-lattice transitions occur.
A classification of these states can be obtained from computing the central charge $c$ or certain order parameters for the vortex lattice (namely the vortex density $\rho_v$ or the average value of local
currents $j_{\rm R}$ on rungs) or the density imbalance $\Delta n$ between the two legs that is nonzero in the BLP phase.
We also list the number $q$ of elementary four-site plaquettes that the unit cell in a given phase contains.

The plan of the paper is the following. We first introduce the model in two different gauge conventions as well as some key observables in Sec.~\ref{sec:model}.
Details on our main method, the DMRG technique, are provided in Sec.~\ref{sec:dmrg}. 
Section~\ref{sec:vortexlattices} discusses the various vortex lattices that are stable at low particle densities as well as the transitions into neighboring phases.
In Sec.~\ref{sec:reversal}, we provide an intuitive explanation of the chiral-current reversal that is tied to certain vortex lattices and study the 
effect of temperature. Section~\ref{sec:CDW} is devoted to the CDW$_{1/4}$ state that exists at filling $\rho=1/4$. The BLP state is studied in Sec.~\ref{sec:BLP},
both analytically and numerically. We summarize our findings in Sec.~\ref{sec:sum}. An Appendix contains our results for additional incommensurabilities in the Meissner phase and a discussion of their possible interpretation.

\section{Model and chiral current}
\label{sec:model}

The system is described by the following Hamiltonian
\begin{align}
H_{\rm rung} = &-J \sum_{r} ( b^\dagger_{1,r} b_{1,r+1} + b^\dagger_{2,r} b_{2,r+1}) \nonumber\\
&- J_\perp \sum_{r}  {\e}^{{\ii} r \phi} b^\dagger_{1,r} b_{2,r} + \mbox{H.c.} \nonumber\\
&+\frac{U}{2} \sum_{r,\ell}  n_{\ell,r} (n_{\ell,r}-1) \;,
\label{eq:ham_rung}
\end{align}
with the matrix elements corresponding to hopping along the rungs and legs of the ladder $J_\perp$ and  $J$, respectively. $U$ is the strength of the onsite interaction (we  consider repulsive interactions
in this work unless stated otherwise). $b^\dagger_{\ell,r}$ creates a particle in the $r$-th site on the leg $\ell=1,2$ and $n_{\ell,r} = b^\dagger_{\ell,r} b_{\ell,r}$.
The total number of bosons is denoted by $N$, while the number of sites in each leg is $L$ (i.e., $1\leq r \leq L$).
We define the particle filling as $\rho=N/(2L)$.  
In the following, we set  $J=1$ as the unit of energy $(\hbar=1)$. 

The model exhibits a gauge  freedom in  choosing different distributions of the Peierls phases as long as the total flux per ladder plaquette remains invariant. For instance,  
by means of a unitary transformation to new bosonic operators
\begin{equation}
\label{ut}
\tilde b_{1,r}={\e}^{-{\ii} r \frac{\phi}{2}}  b_{1,r},\quad \tilde b_{2,r}={\e}^{{\ii} r \frac{\phi}{2}}  b_{2,r}
\end{equation}
we can make the hopping matrix elements along the rungs real, but instead the hopping matrix elements along the legs become complex. The  Hamiltonian~\eqref{eq:ham_rung} is then given by
\begin{align}
H_{\rm leg} = &- \sum_{r} ( {\e}^{{\ii}\frac{\phi}{2}} \tilde b^\dagger_{1,r} \tilde b_{1,r+1} + {\e}^{-\ii\frac{\phi}{2}} \tilde b^\dagger_{2,r} \tilde b_{2,r+1}) \nonumber\\
&- J_\perp \sum_{r} \tilde b^\dagger_{1,r} \tilde b_{2,r} + \mbox{H.c.} \nonumber\\
&+ \frac{U}{2} \sum_{\ell,r} \tilde n_{\ell,r} (\tilde n_{\ell,r}-1) \;.
\label{eq:ham_leg}
\end{align}
We shall refer to these two gauge choices described by the Hamiltonians Eqs.~\eqref{eq:ham_rung} and \eqref{eq:ham_leg} as {\it rung gauge} and {\it leg gauge}, respectively.

An important, experimentally accessible~\cite{Atala2014,Mancini2015,Stuhl2015} observable in ladder systems is the local current, defined on either bonds in the bulk of the system or on the boundaries. 
From the continuity equation
$\left\langle \frac{dn_{{\bf r}}}{dt}\right \rangle = {\rm i}\left\langle [H, n_{\bf r}]\right\rangle =-\sum_{\left\langle {\bf s}\right\rangle} j({\bf r}\to {\bf s})$
we can define the current $j({\bf r}\to {\bf s})$ from a site ${\mathbf r}$ to a neighboring site ${\mathbf s}$ (where ${\mathbf r}=(\ell,r)$). 
In particular, for the model \eqref{eq:ham_rung}, we 
obtain the local currents on legs and rungs from 
\begin{align}
j^{\parallel}_{\ell,r} &= i a \left( b^\dagger_{{\ell,r+1}} b_{\ell,r} - b^\dagger_{\ell,r} b_{{\ell,r+1}} \right)
\nonumber\\
j^{\bot}_{r} &= i J_{\bot} a \left( e^{-i r \phi} b^\dagger_{1,r} b_{2,r} - e^{i r \phi} b^\dagger_{2,r} b_{1,r} \right)\,.
\end{align}

Apart from the configuration of local currents, the average current that circulates along the boundary of the system may reveal important properties of the quantum phases. 
This so-called chiral current (also dubbed edge, screening or Meissner current) is defined as
\begin{align}
j_c =\frac{1}{N}\sum_{r} \langle j^{\parallel}_{1,r}-j^{\parallel}_{2,r} \rangle\,.
\end{align}
For a two-leg ladder, we may obtain $j_c$ from the Hellmann-Feynman theorem as the derivative of the ground-state energy $E_0$ per particle with respect to the flux $\phi$
\begin{align}
j_c = \partial_\phi E_0 / N \;.
\label{eq:fl_jc_hellmann}
\end{align}
This, in particular, shows that the expectation value of the  chiral current is gauge invariant.
The experimental realizations of ladder models with either superlattices ~\cite{Atala2014} or synthetic lattice dimensions \cite{Mancini2015,Stuhl2015} 
correspond to the gauge choice  with complex hopping matrix elements  along the rungs as in Eq.~(\ref{eq:ham_rung}).
The superlattice experiment \cite{Atala2014} used flux values of $\phi\simeq \pm \pi/2$, 
while the synthetic lattice dimension experiments were operated at $\phi \simeq \pm 2\pi/3$ \cite{Stuhl2015}  and
 $\phi\simeq  \pm 0.37 \pi$~\cite{Mancini2015}.

\section{Numerical method: Density matrix renormalization group technique}
\label{sec:dmrg}

Most results of this work are obtained by means of large
scale numerical density matrix renormalization group 
\cite{White1992, Schollwoeck2005,Schollwoeck2011} simulations~(DMRG), which is a standard method 
for the simulation of one-dimensional chains or ladder-like systems at zero temperature.
We simulate the ladder model of Eq.~\eqref{eq:ham_leg} with up to $L=160$ rungs, typically using 1000 DMRG states. 
We control the accuracy by enforcing a sufficiently small discarded weight.

The repulsive interactions allow us to employ a cutoff for the  occupation of bosons per site to address large system sizes.
We typically use a cutoff of $n_{\rm max}=4$ bosons for $U\gtrsim J$, $n_{\rm max}=3$ for $U\gtrsim10J$ and $n_{\rm max}=2$ for $U\gtrsim30 J$ and fillings $\rho<1$. By comparison with larger and smaller cutoffs we have ensured the independence of the numerical data on the cutoff, for the quantities shown in this work. A detailed analysis of the 
dependence on $n_{\rm max}$ is contained in the supplemental material of  Ref.~\cite{Greschner2015}.

Close to the V-SF to VL$_{\rho_v}$-SF boundaries, the DMRG simulations tend to converge
to metastable excited states with a varying  vortex density. To overcome this problem, we  perform several calculations (sometimes, up to sixteen runs) starting from different randomly chosen
initial states. Selecting those states with the lowest energy  gives the piecewise continuous results  for, e.g.,  the chiral current $j_c$ shown in Fig.~\ref{fig:VL_0.25}.

\section{Vortex lattices}
\label{sec:vortexlattices}

The existence of vortex lattices in the bosonic two-leg ladder in the presence of a uniform 
gauge field was initially predicted from the consideration of large-capacitance Josephson-junction arrays in the so-called  classical limit \cite{Kardar1986}. Since vortex lattices break a discrete symmetry of the model, they are robust to finite quantum fluctuations, as confirmed by a bosonization analysis of the $J_\perp \ll J $ regime \cite{Orignac2001}.
For the strongly interacting, low-density regime and arbitrary $J_\perp/J$, vortex lattices were first seen for filling $\rho=1$ and $\phi=\pi$ \cite{Dhar2012, Dhar2013}. In those  studies \cite{Dhar2012, Dhar2013}, the emphasis was put on the spontaneous breaking of time-reversal symmetry and thus this state was dubbed a 
chiral Mott insulator \cite{Dhar2012, Dhar2013} (see also \cite{Zaletel2014} for a discussion of chiral Mott insulators in two dimensions). In this state, 
translation symmetry is also broken spontaneously, which is, however, not independent from time-reversal symmetry breaking, since translation with respect to one ladder plaquette accompanied with time reversal remains intact. Hence, one can talk about breaking of translation symmetry instead of time-reversal symmetry in this state, interpreting it as a usual vortex lattice state packed with the maximal number of vortices (siting on every other plaquette) \cite{Greschner2015}.

 Our DMRG study of the hard-core boson limit $U/J=\infty$ did not observe any vortex lattice (at any density $\rho$) but merely Meissner and vortex-liquid states \cite{Piraud2015}.
Vortex-lattice states are, at low densities $\rho \leq 1$, stable at intermediate interactions as we demonstrated in Ref.~\cite{Greschner2015}:
there, we reported evidence for the existence of vortex lattices at $\rho_v=1/2$ and $\rho_v=1/3$ for interaction strength $ 1 \lesssim U/J \leq 10$.
While detailed phase diagrams for such  intermediate values of $U/J$ and low densities have been reported and discussed in Ref.~\cite{Greschner2015}, we 
here focus on the properties of vortex lattices at particle fillings $\rho=0.8$ and $\rho=0.5$ and the transitions between these vortex lattices and other quantum phases (see Secs.~\ref{sec:VL12} and \ref{sec:VL13}).
Moreover, we have detected another stable vortex lattice at $\rho_v=1/4$ (see Sec.~\ref{sec:VL14}). This vortex lattice is interesting since with such a large unit cell, more information on the location of vortices and their extension can be extracted.

In our discussion, we devote particular attention to various measures of the vortex density $\rho_v$. For instance, one can extract $\rho_v$
from the Fourier transform of rung-current patterns (in the case of open boundary conditions), the momentum distribution function (at least in the leg gauge as defined above),
or from modulations in the particle density (see Secs.~\ref{sec:vl_over} and \ref{sec:mdf_rhov}). While these three measures yield consistent results in vortex lattices and (most) vortex liquids, the outcome can differ in, for instance, the Meissner phase for certain parameter regimes. Finally, in Sec.~\ref{sec:VLsyntheticdimensions} we discuss the stability of vortex lattices against augmenting the Hamiltonian with nearest-neighbor repulsive interactions on the rungs (as appropriate for synthetic-lattice dimension realizations). 

\subsection{Overview: quantum phases and vortex lattices at density $\rho=0.8$}
\label{sec:vl_over}

\begin{figure}[tb]
\includegraphics[width=1\linewidth]{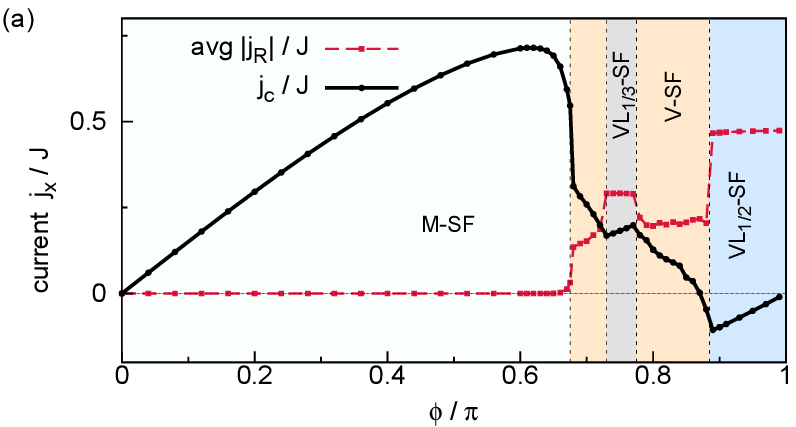}
\includegraphics[width=1\linewidth]{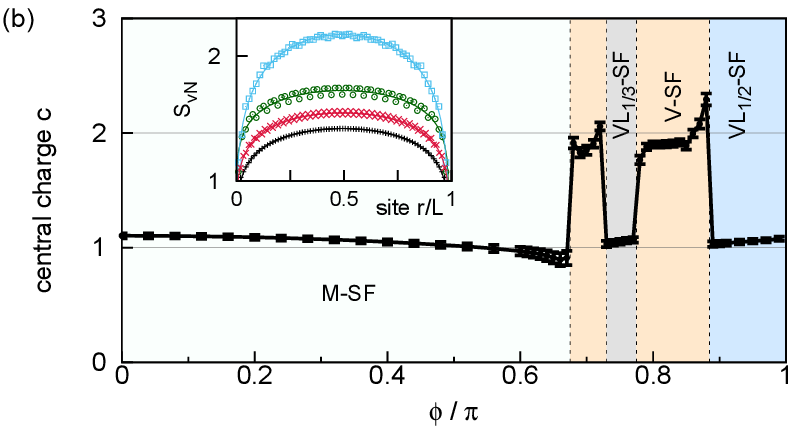}
\includegraphics[width=1\linewidth]{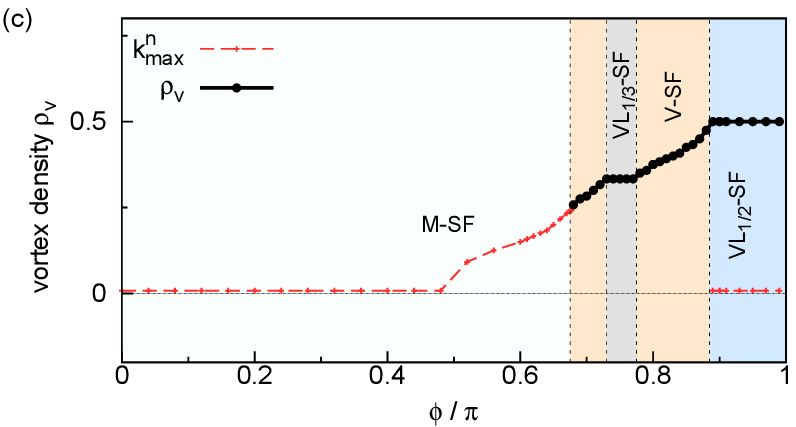}
\caption{Sequence of quantum phases at $\rho=0.8$, $J_\perp=1.6 J$, $U=2J$ as a function of flux $\phi$ probed by several measures:
(a) Chiral current $j_c$ and average rung current avg $|j_{\rm R}|$, (b) central charge $c$ and (c) estimates 
of vortex density $\rho_v$. The vortex density is computed from the Fourier transform of the rung current pattern on systems with open boundary conditions using
Eq.~\eqref{eq:rhov} or from analyzing local density fluctuations and plotting the wavenumber $k_{\rm  max}^n$ of the maximum in the corresponding Fourier transformation 
(see Sec.~\ref{sec:vl_over}). 
$k^n_{\rm max}$ denotes the position of the  maximum in the Fourier transform of the local density fluctuations (see the text). 
The inset in (b) shows the block-entanglement entropy $S_{\rm vN}$ for (from top to bottom) $\phi/\pi=0.8$ (V-SF), $0.74$ (VL$_{1/3}$-SF), $0.62$ (M-SF) and $0.96$ (VL$_{1/2}$-SF).}
\label{fig:cut_n0.5_p0.9}
\end{figure}

Figure~\ref{fig:cut_n0.5_p0.9} depicts various observables for the characterization of vortex-lattice and vortex-liquid phases in the model given in Eq.~\eqref{eq:ham_rung} as a function of flux $\phi$. The numerical results were obtained from DMRG simulations for $U=2J$ and a low filling $\rho=0.8$. A sequence of phases starting from the M-SF, a first vortex-liquid, the VL$_{1/3}$-SF, another sliver of the V-SF, and finally, the VL$_{1/2}$-SF is realized for these parameters. 

In the Meissner and vortex-lattice phases, the chiral current $j_c$ exhibits a characteristic quasi-linear increase with the flux $\phi$ as shown in Fig.~\ref{fig:cut_n0.5_p0.9}(a). The average rung current (also plotted in Fig.~\ref{fig:cut_n0.5_p0.9}(a)) 
\begin{align}
{\rm avg} |j_R| = \frac{2}{L} \sum_{r=-L/4, \dots, L/4} |j^{\bot}_{r}|
\label{eq:avgjr}
\end{align}
exhibits a stable large plateau in the vortex-lattice phases. The transition from the vortex-lattice phases to the V-SF phases is indicated by a marked drop of ${\rm avg} |j_R|$.

Apart from identifying the vortex-lattice and Meissner phases by their characteristic local current configurations (see the discussion in Ref.~\cite{Greschner2015} and Fig.~\ref{fig:patterns}), they may  clearly be  
discriminated from the vortex-liquid phase by calculating the central charge $c$, which may be extracted from scaling properties of the entanglement entropy $S_{\rm vN}(l)$ of a a subsystem of length $l$ embedded in a chain of a finite length $L$~\cite{Vidal2003,Calabrese2004}
\begin{align}
S_{\rm vN}(l) = \frac{c}{6} \log\left[ \frac{L}{\pi} \sin \frac{\pi l}{L} \right] + \cdots \,,
\label{eq:ee_obc}
\end{align}
where we have omitted non-universal constants and higher-order oscillatory terms due to the finite size of the system.
For a more detailed discussion of the behavior of the entanglement entropy in this model and the extraction of $c$, see the supplemental material of \cite{Piraud2015,Greschner2015}. 
In praxis, we compute $S_{\rm vN}$ for blocks that contain $r$ rungs, i.e., we discard blocks that would cut a rung and thus 
we plot $S_{\rm vN}$ versus the number of rungs $r$ in the block in the figures.

In Fig.~\ref{fig:cut_n0.5_p0.9}~(b), we depict the extracted central charge from fitting  Eq.~\eqref{eq:ee_obc} to the numerical data, 
which is well consistent with $c=1$ in the vortex lattices atop the superfluid phase (such as the VL$_{1/2}$-SF and VL$_{1/3}$-SF states) 
and the M-SF phase and $c=2$ in the V-SF phase. Interestingly, 
as shown in the inset of Fig.~\ref{fig:cut_n0.5_p0.9}(b), the entanglement entropy in systems with open boundaries 
 exhibits small oscillations that follow the lattice structure of the vortex-lattice phases.

We next estimate the vortex density $\rho_v$ by analyzing the rung-current configurations $\langle j_{r}^{\perp}\rangle$. We follow our previous analysis detailed in 
\cite{Piraud2015},  
where we introduced a measure for the vortex density, given by the inverse typical distance between the vortex cores $l_v$ 
\begin{equation}
\rho_v=l_v^{-1} \, .\label{eq:rhov}
\end{equation}
We extract this distance from the Fourier transform of the real-space patterns of the rung currents $\langle j_{r}^{\perp}\rangle$ (which in the vortex fluids are obviously discernible due to finite-size effects for open boundary conditions). 
The resulting vortex density depicted in Fig.~\ref{fig:cut_n0.5_p0.9}(c) shows the typical 
devil's staircase-like structure predicted in Ref.~\cite{Kardar1986,Orignac2001}, i.e., a sequence of the Meissner phase ($\rho_v=0$), incommensurate vortex liquids and vortex lattices. We draw the reader's attention to the sharp jump of the vortex density at the boundary of the M-SF to the neighboring vortex-liquid phase located in the vicinity of $\phi=0.7\pi$. This jump may be indicative of a first-order transition.

Alternatively, the vortex density may  be related to local density fluctuations in systems with open boundaries (as already discussed by Ref.~\cite{Atala2014} for noninteracting particles). The Fourier transform of such local density fluctuations exhibits a sharp peak at a momentum $k^n_{\rm max}$, which, outside of the VL$_{1/2}$-SF and Meissner phases coincides with $\rho_v$ (see Fig.~\ref{fig:cut_n0.5_p0.9}(c)). 
In the VL$_{1/2}$-SF, there are no density fluctuations by symmetry \cite{Greschner2015} and hence $k^n_{\rm max}=0$. 
Interestingly, in the Meissner phase and for a finite flux,
 we also observe finite, small-amplitude fluctuations in the density as a precursor of the transition to the vortex phases.
As we shall see below in Sec.~\ref{sec:mdf_rhov}, in the superfluid phase, where single-particle correlation functions decay algebraically, we can also
 estimate the vortex density from the momentum distribution function. We will continue our discussion of the vortex density in Sec.~\ref{sec:mdf_rhov}.

\subsection{Momentum distribution function: Experimental observable and  a possible  measure of vortex density}
\label{sec:mdf_rhov}

\begin{figure*}[tb]
\includegraphics[width=0.47\linewidth]{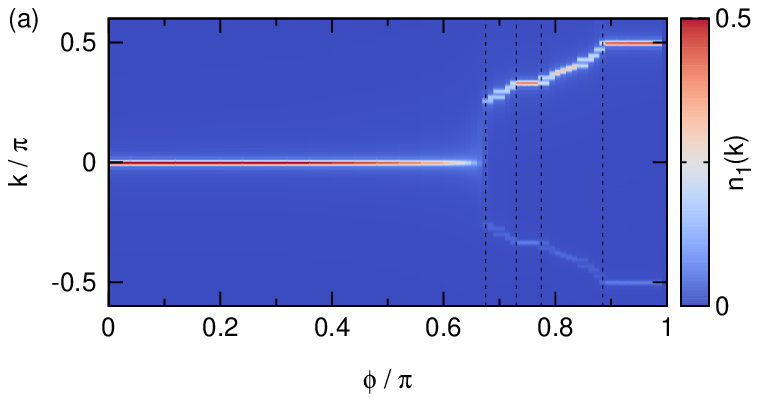}
\hspace{0.2cm}
\includegraphics[width=0.47\linewidth]{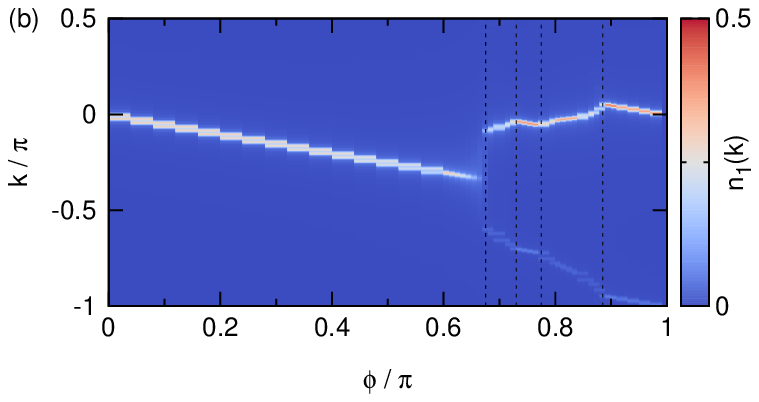}
\caption{Momentum distribution for $\rho=0.8$, $J_\perp=1.6 J$, $U=2J$ for (a) the leg gauge (see Eq.~\eqref{eq:ham_leg}) and (b) the rung gauge (see Eq.~\eqref{eq:ham_rung}). 
Note that (a) and (b) are related by a linear shift by $\phi/2$ due to the exact gauge transformation of Eq.~\eqref{peakmomentum}.}
	\label{fig:cut_mom_phi}
\end{figure*}

In the following, we study the momentum distributions along the legs of the ladder $n_{\ell}(k)$, with $\ell=1,2$:
\begin{align}
n_\ell(k) =\frac{1}{L} \sum_{r,r'} {\e}^{{\ii} k (r-r')} \langle b_{\ell,r}^\dagger b_{\ell,r'} \rangle \;,
\end{align}
which are measurable in time-of-flight experiments \cite{Bloch2008}. 
Using Eq.~(\ref{ut}) one can see that the momentum distributions in the two different gauges defined in Sec.~\ref{sec:model} are related to each other via 
\begin{equation}
n_1(k)=\tilde n_1(k- \phi/2),\quad n_2(k)=\tilde n_2(k+ \phi/2).
\end{equation}
Figure~\ref{fig:cut_mom_phi}(a) shows that, in the leg gauge, the central peak of $n_{\ell}(k)$ perfectly coincides with the 
vortex density $\rho_v$ of the system (compare Fig.~\ref{fig:cut_n0.5_p0.9}(c)). For the rung gauge, 
the peak position exhibits an additional shift and behaves qualitatively similarly to the chiral current $j_c$ (compare Fig.~\ref{fig:cut_n0.5_p0.9}(a)). 

We next compute the momentum distributions in Meissner, vortex-lattice and vortex-liquid phases with the help of an effective field-theory approach based on bosonization.
We introduce two pairs of bosonic fields ($\theta_{\ell},\phi_{\ell}$), describing the  phase and density fluctuations of bosons on leg $\ell$, respectively, 
with $[\theta_{\ell}(x),\partial_y\phi_{\ell'}(y)]=i\delta_{\ell,\ell'}\delta(x-y)$. The low-energy properties of the model Eq.~(\ref{eq:ham_rung}) are governed by the following Hamiltonian density
\begin{eqnarray}
{\mathcal H}&=&\frac{ v_{+}}{2}\left[ \frac{(\partial_x \phi_{+})^2}{ K_{+}}+ K_{+}(\partial_x \theta_{+})^2  \right] \label{eq:hdens} \\
&+&\frac{ v_{-}}{2}\left[ \frac{(\partial_x \phi_{-})^2}{ K_{-}}+ K_{-}(\partial_x \theta_{-}+\frac{\phi}{\sqrt{2\pi}})^2  \right]  \nonumber\\
&-&\!\!\!\!\sum_{q=1,2,...} \!\! \!\! \cos { \sqrt{2\pi} q\theta_{-}} \!\!\! \sum_{m=0,1,...} \!\!\! \!\!\lambda_{q,m}\cos{[m\sqrt{8\pi} \phi_{+} \!+4m\pi nx]}\nonumber
\end{eqnarray}
where $\phi_\pm=(\phi_1\pm \phi_2)/\sqrt{2}$, $\theta_+=(\theta_1+ \theta_2)/\sqrt{2}$. The expression for $\theta_-$ depends on the gauge:  
for the case where the Peierls phases are along the  rungs, $\theta_-=(\theta_1- \theta_2-\phi x/\sqrt{\pi})/\sqrt{2}$, whereas for the gauge where the Peierls phases are along the  legs,  $\theta_-=(\theta_1- \theta_2)/\sqrt{2}$. $K_{\pm}$ are Luttinger-liquid parameters 
corresponding to the total and relative fluctuations on the two-leg ladder and $v_{\pm}$ are the corresponding velocities, 
which in general need to be determined from a comparison with numerics. 

Since the flux couples to the topological charge of the sine-Gordon model describing the antisymmetric sector, we explicitly separate out 
the zero-momentum mode in the field expansion which is related to the vortex density \cite{Kardar1986}
\begin{equation}
\theta_-(x)= -\sqrt{2\pi} \rho_v x+\frac{1}{\sqrt{L}}\sum_{p\neq 0}e^{ipx}\theta_p\,.
\end{equation}
For small values of $\phi$, the most important term in Eq.~(\ref{eq:hdens}) is the one proportional to $\lambda_{1,0}\sim J_{\bot}$, which 
at any filling and $U$ opens a gap  in the antisymmetric sector already for an arbitrarily small interchain tunneling  and pins $\langle \theta_-\rangle$ (i.e., it 
locks the relative phase of bosons on the two legs). Thus, the system is in the Meissner phase as long as $\phi<\phi_c$, where $\phi_c$ is a soliton gap of the quantum sine-Gordon model describing the antisymmetric sector \cite{Kardar1986}.

With increasing flux, the terms proportional to $\lambda_{q,0}$ with $q>1$ can become commensurate (for integer values of $q \rho_v$) and relevant (for $q^2 \le 4K_-$) in vortex-lattice states with a $q$-fold degenerate ground state and a spontaneously enlarged unit cell with  $q$ plaquettes \cite{Orignac2001}. 

Using the representation
\begin{equation}
b_{1(2)}\to  e^{i\sqrt{\pi}(\theta_+\pm \theta_-)/\sqrt{2}}
\end{equation}
(where the $+(-)$ sign corresponds to $\ell=1(2)$)
we can calculate the momentum distributions, say along the first leg of the ladder in vortex-lattice phases (the expression also applies to the  Meissner phase where $\rho_v=0$), resulting in 
\begin{equation}
\label{mdvl}
n_1(k)\sim {|k-k_p|^{\frac{1}{4K_+}-1}}.
\end{equation}
On the other hand, in the vortex-liquid phase, where the antisymmetric sector is also described by the Luttinger liquid, we obtain 
\begin{equation}
\label{mdv}
n_{1}(k)\sim {|k - k_p|^{\frac{K_++K_-}{4K_+K_-}-1}}\,.
\end{equation}
The position of the peak in the momentum distribution is gauge dependent
\begin{equation}
\label{peakmomentum}
k_p=-\pi \rho_v+ \alpha \phi/2,
\end{equation}
where $\alpha=1$ for the rung gauge  and $\alpha=0$ for the leg gauge.
Since the Luttinger-liquid parameters $K_{\pm}$ are positive numbers, one can see by comparing Eqs.~(\ref{mdvl}) and (\ref{mdv}) 
that in vortex-lattice states, the momentum distribution has a larger weight at its peak value than in vortex-liquid states, which is confirmed by the
numerical data shown in Fig.~\ref{fig:cut_mom_phi}.

In the rung gauge (the case relevant for experiments), 
the kinematic and canonical momenta of particles coincide and hence the position of the peak in the momentum distribution function is related to the chiral current (all bosons on leg $1 (2)$ can be thought of as being quasi-condensed at the momentum $k_p(-k_p)$). 
In the weak-coupling limit we can hence use the approximation
\begin{equation}
\label{ccmd}
j_c \sim \sin{(k_p)}- \sin{(-k{_p})} =2\sin{(\phi/2 -\pi \rho_v)},
\end{equation}
where the sin functions appear due to the presence of the lattice and the fact that the currents are related to particle velocities rather than to their quasi-momenta. 

From  Eq.~(\ref{ccmd}), one can also see that the chiral current reverses its circulation direction in certain VL$_q$ states and, in particular, in VL$_{1/2}$ states. 
The chiral current can change its sign in other vortex lattices as well.
For instance,  if a VL$_{1/3}$ state is realized for flux values including the point $\phi= 2\pi/3$ then, from Eq.~(\ref{ccmd}), we infer that  a sign change of $j_c$ occurs at $\phi=2\pi/3$, where $j_c$ vanishes. A similar conclusion has been reached previously by Orignac and Giamarchi \cite{Orignac2001}. The physical mechanism underlying the chiral-current reversal is a spontaneous increase of the effective flux piercing the unit cell of vortex-lattice states, as discussed in our recent work \cite{Greschner2015}.

\subsection{Vortex lattice at $\rho_v=1/2$ and density $\rho=1/2$}
\label{sec:VL12}

\begin{figure}[tb]
\includegraphics[width=1\linewidth]{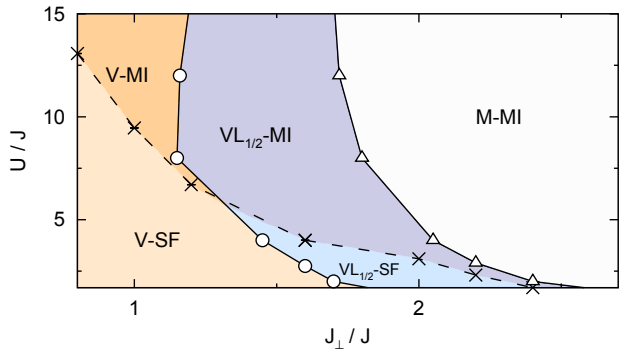}
\caption{Phase diagram for $\rho=0.5$ and $\phi=0.9\pi$ in the $U/J$ versus $J_\perp$ plane. 
Symbols denote estimated points of the vortex-fluid-to-vortex lattice  ($\circ$), the VL$_{1/2}$-MI-to-M-MI ($\Delta$) and the MI-SF ($\times$) phase transitions (see Sec.~\ref{sec:VL12}). Straight lines and shadings are  guides to the eye.}
\label{fig:pd_U_Jperp_n0.5_p0.9}
\end{figure}

In the following we study the vortex-lattice phase for $\rho_v=1/2$ at filling $\rho=1/2$. Figure~\ref{fig:pd_U_Jperp_n0.5_p0.9} shows 
the phase diagram as a function of $J_\perp/J$ and $U/J$. While in the limit of hard-core bosons $U/J\to \infty$, 
there is just a direct transition from a V-MI to a M-MI phase \cite{Piraud2015}, for finite values $U/J$, an intermediate VL$_{1/2}$-MI phase exists.

\begin{figure}[tb]
\includegraphics[width=1\linewidth]{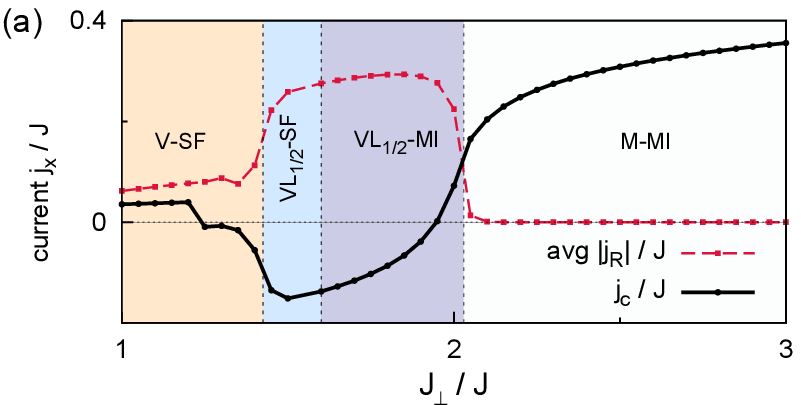}
\includegraphics[width=1\linewidth]{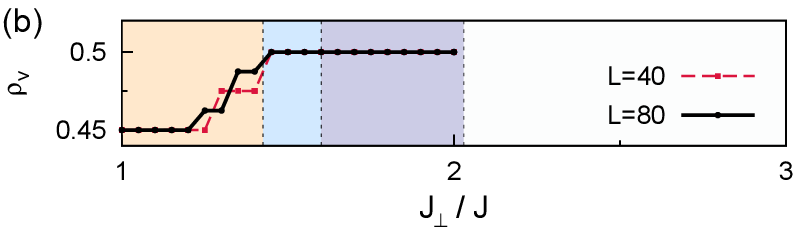}
\caption{Quantum phases at $\rho=0.5$, $\phi=0.9\pi$: Cut through the phase diagram Fig.~\ref{fig:pd_U_Jperp_n0.5_p0.9} at  $U/J=4$. 
(a) Chiral current $j_c/J$ and average rung-current ${\rm avg} |j_R|$; (b) vortex density $\rho_v$, all versus $J_\perp/J$. 
Dashed lines denote the positions of the phase transitions.}
\label{fig:cut_U_Jperp_n0.5_p0.9_U4.0}
\end{figure}

First indicators of the vortex-liquid (V-SF) to VL$_{1/2}$-SF transition can be detected in the behavior of $j_c$ and ${\rm avg} |j_R|$. Namely, there is a  marked increase of ${\rm avg} |j_R|$ and a kink in the chiral current $j_c$ (see Fig.~\ref{fig:cut_U_Jperp_n0.5_p0.9_U4.0}(a)). 
Vortex lattices break translational symmetry and hence the finite values of local rung currents are their true thermodynamic feature (open boundaries select one of the degenerate ground states), whereas nonzero values of local rung currents that we observe in vortex-liquid states are caused by the combined effect of open boundaries and finite system size. Hence, upon increasing the system size the jump in ${\rm avg} |j_R|$, when transitioning from the vortex-liquid to the vortex-lattice state, becomes more and more pronounced.
A more precise estimate of the phase transition point is possible via the calculation of the vortex density $\rho_v$. The results are  
illustrated in Fig.~\ref{fig:cut_U_Jperp_n0.5_p0.9_U4.0}(b). 
For small values of $U/J$ such as $U/J=4$ (see Fig.~\ref{fig:cut_U_Jperp_n0.5_p0.9_U4.0}(a)), the vortex-lattice phases VL$_{1/2}$-SF and VL$_{1/2}$-MI give rise to an inversion of the sign of the chiral current $j_c<0$ as discussed in Ref.~\cite{Greschner2015}.

\begin{figure}[b]
\includegraphics[width=1\linewidth]{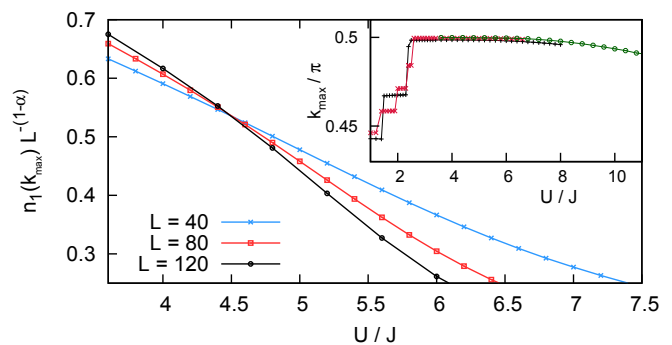}
\caption{Scaling of the peak of the momentum distribution $n_1(k_{\rm max})$ as a function of $U/J$ close to the BKT transition from the VL$_{1/2}$-SF to the VL$_{1/2}$-MI phase ($J_\perp=1.6 J$, $\rho=0.5$, $\phi=0.9\pi$). For the exponent,
 we choose $\alpha=1/4$ corresponding to the expected scaling of the single-particle correlation functions. The crossing point 
of all curves marks the BKT-transition point. The inset shows $k_{\rm max}$ over a larger range of $U/J$ including the V-SF to VL$_{1/2}$-SF transition where $k_{\rm max}=\pi/2$. 
}
\label{fig:momscaling_n0.5}
\end{figure}

The transitions from the SF to MI phases are of the Berezinskii-Kosterlitz-Thouless (BKT) type and their position is the point where the Luttinger-liquid parameter $K_\rho$ takes the value $K_\rho=1$.
We extract $K_\rho$ from the long-wavelength behavior of the static structure factor, $ \frac{1}{L} \sum_{i,j} e^{{\rm i} (i-j) k} \langle n_{i} n_{j}\rangle$~\cite{Moreno2011}.
For the VL$_{1/2}$-SF to VL$_{1/2}$-MI phases we may verify this estimate by an analysis of single-particle correlation functions, which are predicted to decay as $\langle a^{\dagger}_{l,r}a_{l,r+x} \rangle  \sim x^{-\frac{1}{4}}$ at the transition. 
We study the finite-size scaling behavior of peaks in the quasi-momentum distribution function~\cite{Greschner2015} 
for extracting the BKT-transition point. An example is illustrated in Fig.~\ref{fig:momscaling_n0.5}.

\begin{figure}[b]
\includegraphics[width=1\linewidth]{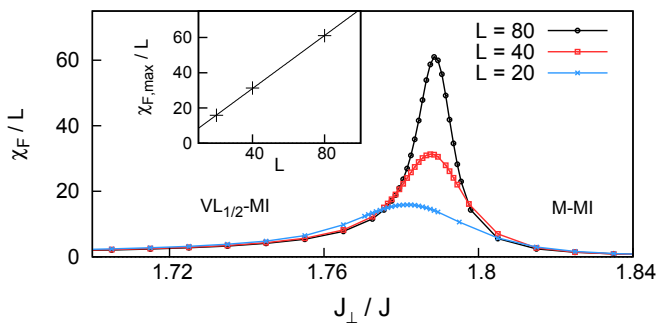}
\caption{Fidelity susceptibility close to the VL$_{1/2}$-MI to M-MI transition for $\rho=0.5$, $\phi=0.9\pi$, and $U/J=8$, showing the Ising character of the transition.}
\label{fig:cut_U_Jperp_n0.5_p0.9_U8.0_fs}
\end{figure}

For large values of $J_\perp/J$, we observe a continuous Ising-type phase transition from the VL$_{1/2}$-MI to the M-MI phase. In order to examine this phase transition, we calculate the  ground-state fidelity susceptibility $\chi_{FS}$~\cite{Gu2010}
\begin{eqnarray*} 
\chi_{FS}(J_\perp)&=& \lim_{\delta J_\perp\to 0} \frac{-2 \ln |F| }{(\delta J_\perp)^2}
\end{eqnarray*}
from the overlap of the ground-state wave functions
$F = \langle \Psi_0(J_\perp) | \Psi_0(J_\perp + \delta J_\perp) \rangle$. Figure~\ref{fig:cut_U_Jperp_n0.5_p0.9_U8.0_fs} depicts the behavior of $\chi_{FS}(J_\perp)$ in the vicinity of the   VL$_{1/2}$-MI to M-MI  transition for several system sizes $L$. The quadratic increase of $\chi_{FS}(J_\perp) \sim L^2$ confirms an Ising-type character~\cite{Venuti2007, Gu2010, Greschner2013b}.

\subsection{Vortex lattice at $\rho_v =1/3$ and $\rho=0.8$}
\label{sec:VL13}

\begin{figure}[tb]
\includegraphics[width=1\linewidth]{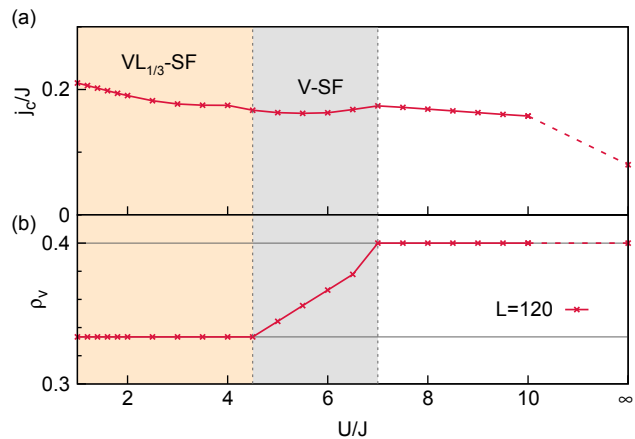}
\caption{Stability of the vortex lattice at $\rho_v=1/3$: (a) Chiral current and (b) vortex density from Eq.~\eqref{eq:rhov} versus $U/J$.
 ($J_\perp=1.6 J$, $\phi/\pi=0.75$, $\rho=0.8$, $L=120$).}
\label{fig:scanU}
\end{figure}

Apart from the vortex lattice at $\rho_v={1/2}$ we have also resolved a  VL$_{1/3}$-SF state in the regime of weak inter-particle interactions but low density in our previous work \cite{Greschner2015}. 
At this point we would like to discuss the fate of the VL$_{1/3}$-SF state as the interaction strength is increased, keeping the value of the flux at  $\phi \simeq 3\pi/4$ 
and filling $\rho=0.8$ fixed. Figure~\ref{fig:scanU} shows the  vortex density $\rho_v$ (obtained from the Fourier transform of the rung currents, Eq.~\eqref{eq:rhov}):
 it is first constant and pinned at $\rho_v=1/3$, as expected for a VL$_{1/3}$ state, then it increases once the system enters into a  vortex-liquid state at $U/J\sim 4.5$.
Thus, this vortex lattice survives up to intermediate values of $U/J$ only, at least for the selected parameters.
Surprisingly, $\rho_v$ becomes flat again at $\rho_v=2/5$ and remains constant even up to $U/J=\infty$.

As another measure of the stability of the VL$_{1/3}$-SF state, we monitor the dependence of the central charge $c$ on $U/J$.
Inside the VL$_{1/3}$-SF state, $c=1$, while it increases to  $c=2$ in the neighboring V-SF at $U/J \gtrsim 4.5$.
Curiously, the central charge drops to $c=1$ again for large values of $U/J>7$. 
The nature of this $c=1$ phase is discussed in the next section.

\subsection{Commensurate vortex structure at $\rho_v=2/5$: vortex lattice, Meissner, or Laughlin-like state?}
\label{sec:VL25}

\begin{figure}[tb]
\includegraphics[width=1\linewidth]{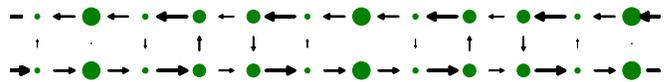}
\caption{Current pattern and density modulations for the hard-core boson limit of Fig.~\ref{fig:scanU} ($J_\perp=1.6 J$, $\phi/\pi=0.75$, $\rho=0.8$, $L=120$). The length and width of the arrows are proportional to the local currents and the radii of the green dots to $\langle n_{\ell,r}-0.75 \rangle $ so as to highlight the density modulations.}
\label{fig:scanU1}
\end{figure}

The data shown in Fig.~\ref{fig:scanU} suggest the existence of another state with a commensurate structure in the rung currents.
An obvious candidate state would be a vortex lattice at $\rho_v=2/5$. We show the pattern of local currents and the density modulations in 
Fig.~\ref{fig:scanU1}.
The finite-size ground-state configuration is seemingly periodic and thus resembles the structure of a vortex lattice.
However, a more detailed analysis suggests that this is not the case.
First of all, based on arguments from bosonization, the stabilization of a VL$_{2/5}$ state at large values of $U/J$ where vortex lattices 
with smaller periodicities have already melted into vortex liquids is extremely unlikely. 

To further elucidate this case we carry out a finite-size analysis of the amplitude of the local density and current modulations for the limit of hard-core bosons 
(see the Appendix) which indicates that the average rung current  vanishes as ${\rm avg} |j_R| \sim 1/L^\alpha$ with $\alpha \approx 0.4 $ in the thermodynamic limit.  
Hence, similar to the vortex-liquid and Meissner states, this state would not break a translational symmetry in the 
thermodynamic limit as local oscillations die out with increasing the system size.
A possible interpretation of this region is a Meissner phase with a Luttinger-liquid parameter $K_\rho<1/2$, which leads to strong correlation effects, a blurred momentum distribution and enhanced density modulations.

There is, however, another possibility that would lead to a commensurate locking of density, namely the one-dimensional analogue of the $\nu=1/2$ Laughlin state (see \cite{Grusdt2014,Petrescu2015} for a discussion). Here, $\nu = N/N_\phi$  is the ratio of particle number over flux quanta.
For this ratio, the results of \cite{Petrescu-aps} may hint at the presence of another commensurate phase in the hard-core boson limit.
In our case, i.e., for the parameters of Fig.~\ref{fig:scanU}, the condition of $\nu=1/2$ is also approximately fulfilled with $\phi/\pi =3/4$ and $\rho=0.8$ (the condition of $\nu=1/2$ would be fulfilled with $\rho=0.8$ and $\phi/(2\pi) =0.4$, since $\rho=0.8$ is equivalent to $\rho=0.2$ for hard-core bosons due to particle-hole symmetry).
While giving a definite answer to this interesting question is beyond the scope of our work,
in the Appendix, we describe a similar situation, which we have encountered studying hard-core bosons close to the boundary between Meissner and vortex-liquid phases for low particle densities~\cite{Piraud2015}, which is also the regime discussed by Petrescu {\it et al.} \cite{Petrescu-aps}.

Finally, the case studied here leads us to another conceptual issue, namely, the {\it  ambiguity} in defining and estimating the {\it vortex density}. First, we note that the vortex density has no direct microscopic definition, unlike particle densities and currents.
As explained in Secs.~\ref{sec:vl_over} and \ref{sec:mdf_rhov}, we extract the vortex density from different quantities, but 
primarily from the  Fourier transform of rung currents from finite-size data with open boundary conditions Eq.~\eqref{eq:rhov} 
or from the position of the  maximum in the momentum distribution function computed in the leg gauge of Eq.~\eqref{eq:ham_leg}. 
In most of the cases considered here and, in particular, in vortex-lattice and most vortex-liquid states, the  vortex densities extracted from these different methods agree with each other. However, this is not necessarily the case in certain other parameter regimes, including the state that we described above at large $U/J>7$, $\rho=0.8$ and $\phi=3/4$. Another example is the Meissner phase at small $U/J=2$ (compare Fig.~\ref{fig:cut_n0.5_p0.9} and its discussion in Sec.~\ref{sec:vl_over}). 
An analogous behavior in the Meissner phase in the hard-core boson limit will be discussed in the Appendix.

\subsection{Vortex lattices with $\rho_v = 1/4$}
\label{sec:VL14}

\begin{figure}[tb]
\includegraphics[width=1\linewidth]{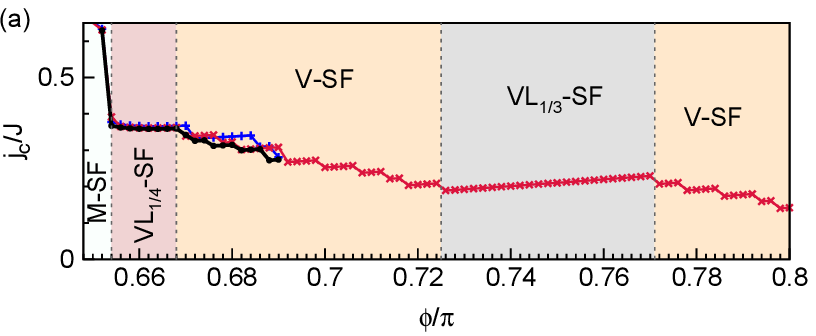}
\includegraphics[width=1\linewidth]{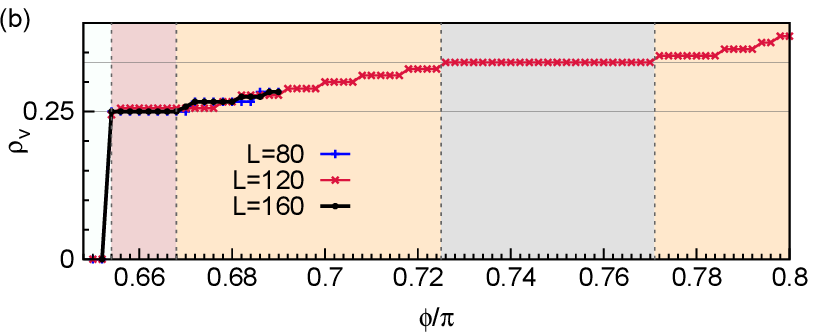}
\includegraphics[width=1\linewidth]{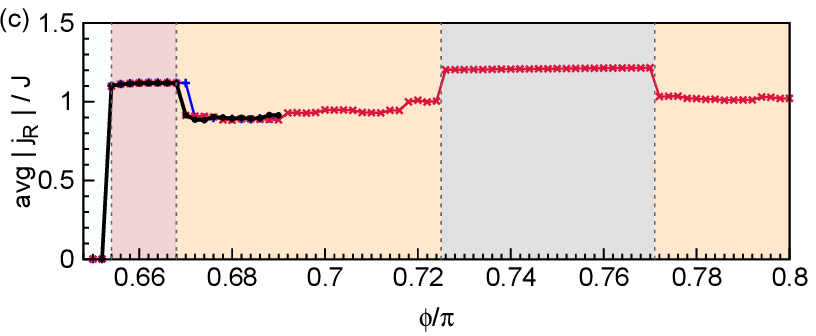}
\caption{Vortex lattice with $\rho_v=1/4$ (VL$_{1/4}$-SF): (a) Chiral current $j_c$, (b) vortex density $\rho_v$ and (c) average rung current 
${\rm avg} |j_R|$ versus flux $\phi$ for $\rho=0.8$, $J_\perp=1.6 J$, $U=J$.}\label{fig:VL_0.25}
\end{figure}

For small values of $U/J$, we resolve another vortex-lattice state at vortex density $\rho_v=1/4$. Figure~\ref{fig:VL_0.25} provides our numerical evidence for the existence of such a vortex lattice. The corresponding configuration of local currents is shown in Fig.~\ref{fig:patterns}(d). 

We can use the example of the VL$_{1/4}$ state to discuss the location and extension of vortices.
In general, the current configurations in VL$_{1/q}$ phases are best interpreted \cite{Greschner2015} as small portions of Meissner regions (extended over $q-1$ consecutive plaquettes and with the 
screening current circulating around the  boundary of the region), 
separated by vortices, which occupy every $q-$th plaquette. 
An analytical estimate of the vortex size $l_0$ can be obtained in the weak-coupling limit and it gives  
 $\l_0\sim \sqrt{J/(2J_{\bot})} a$ \cite{Kardar1986}, which suggests that the vortices are rather tightly localized objects.
Note, however, that in the VL$_{1/4}$-SF state, the rung currents do not fully vanish inside the Meissner portion 
due to the nonzero screening length.

This observation leads us to the interesting question of how the   vortex size compares to  the mean distance between the vortex cores. In our previous analysis of VL$_{1/2}$ and VL$_{1/3}$ states (and vortex fluids in systems with open boundaries) \cite{Piraud2015,Greschner2015}, we usually observe one length scale. However, VL$_{1/2}$ and VL$_{1/3}$ states are the most densely packed vortex lattices and hence are not the optimal cases to resolve the size of individual vortices.  

Moreover, due to symmetry reasons, in the VL$_{1/3}$ states, in the middle rung of each Meissner portion, the rung current vanishes exactly. Such a behavior is expected for any VL$_q$ state with odd $q$, since the Meissner phase of the ladder with an odd number of rungs has a reflection symmetry with respect to the middle rung of the ladder, accompanied with a reversal of the current circulation direction, which implies an exact vanishing of the rung current in the middle rung.

The local particle density shows a strong modulation in the vortex-lattice phases (except for  the vortex lattice at the 
maximal possible vortex density $\rho_v=1/2$). This effect has not been captured in previous bosonization studies~\cite{Orignac2001}.  
Moreover, in the VL$_{1/4}$-SF, we also observe a modulation of the absolute values of the local rung currents, also not captured by previous bosonization analyses.

To summarize, our interpretation of certain plaquettes as the position where the vortex cores are localized in vortex lattices is based on two facts: first, the direction of the local particle currents around theses plaquettes is opposite to the direction of the chiral current circulating around the Meissner region (thus vortices reduce the overall chiral current, consistent with the flux and $J_\perp$ dependence of $j_c$). The second reason is that local particle densities are reduced in the plaquettes where vortices are localized. 
In order to numerically study the size of vortices, we would need access to less closely packed vortex lattices, which is left for future studies.

\subsection{Stability of vortex lattices in systems with a synthetic lattice dimensions}
\label{sec:VLsyntheticdimensions}

For experimental realizations using a synthetic lattice dimension~\cite{Celi2014,Mancini2015, Stuhl2015}, long-range interactions in the rung direction  
have to be taken into account. In our case this amounts to
\begin{align}
H_{V} = V \sum_r n_{1,r} n_{2,r} \;.
\label{eq:ham_Vrung}
\end{align}
The ratio $U/V$ depends on the properties of the atomic species and might be controlled externally by means of Feshbach resonances \cite{Bloch2008} or lattice modulation techniques~\cite{Cardarelli2016}. A reasonable first approximation is to set $U=V$, while here we will allow $V$ to vary between $0<V/J<1$.

\begin{figure}[tb]
\includegraphics[width=1\linewidth]{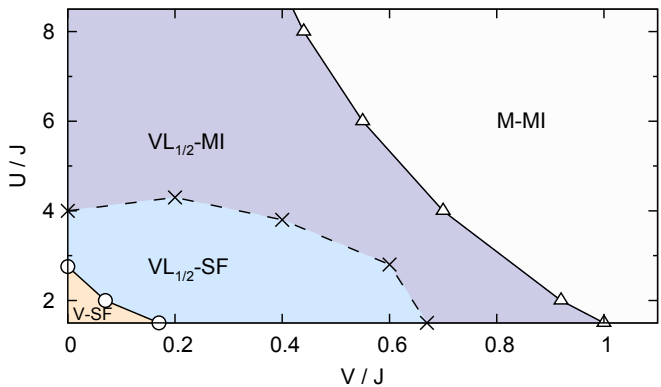}
\caption{Stability of vortex phases against nearest-neighbor interactions on the rungs: Phase diagram for $\rho=0.5$ for $\phi=0.9\pi$, $J_\perp=1.6 J$ as a function of $U/J$ and $V/J$. Symbols denote estimated points of the V-VL ($\circ$), VL-M ($\Delta$) and the MI-SF ($\times$) phase transitions (see the 
discussion in Sec.~\ref{sec:VLsyntheticdimensions}). Straight lines and shadings are  guides to the eye.}
\label{fig:pd_V_U_n0.5_p0.9}
\end{figure}

In Fig.~\ref{fig:pd_V_U_n0.5_p0.9}, we present the phase diagram for finite positive values of $V$, $J_\perp=1.6J$ and $\phi=0.9\pi$. 
The presence of large rung interactions $V/J$ favors the M-MI phases
and suppresses vortex phases, in agreement with the observation made for hard-core bosons on a three-leg ladder \cite{Kolley2015}.

For the parameters of Fig.~\ref{fig:pd_V_U_n0.5_p0.9} and for the relevant case of $U=V$,
 a transition from the VL$_{1/2}$ to the Meissner phase would be expected for $U=V \approx J$. 
However, it is important to note that the regime of stability strongly depends on the particle filling 
since for a low filling, the interaction $V$ becomes less relevant: Anticipating the results of Sec.~\ref{sec:CDW}, 
at quarter filling $\rho=0.25$ a stable VL$_{1/2}$ phase (as well as other interesting phases discussed there) can be found up to large values $U=V\sim 30J$.

\section{Chiral-current reversal: spontaneous vs explicit symmetry breaking}
\label{sec:reversal}

A main result of our previous work \cite{Greschner2015} is the observation of a sign change of the chiral current (i.e., a reversal of its circulation direction) in certain vortex lattice phases. 
We explained this via the mechanism of an increase of the effective flux seen by the particles as a result of the spontaneous breaking of 
lattice translation symmetry in the vortex lattices, which results in a $q$-fold enlarged unit cell. 
If $\phi$ is the flux per plaquette, then the effective flux is $\phi_{\rm eff} = q \phi$ and therefore, the 
chiral current is $j_c = j_c( q \phi)$. Since the chiral current is $2\pi$-periodic, this can correspond to a negative current,
if, for instance, $ \pi <   q\phi < 2\pi$ since $j_c(q\phi) = j_c(q\phi - 2\pi)$.

Among the examples shown in Fig.~\ref{fig:patterns}, the VL$_{1/2}$ states exhibit this behavior, since these vortex lattices are typically stable for $\phi \lesssim \pi$.
Examples for the sign change of the chiral current are shown in Figs.~\ref{fig:cut_n0.5_p0.9}(a) and  \ref{fig:cut_U_Jperp_n0.5_p0.9_U4.0}(a).  

Here, we will provide an intuitive explanation for the effect of the chiral-current reversal in bosonic ladders with a unit cell that is larger than just one plaquette.

\begin{figure}[tb]
\begin{center}
\includegraphics[width=1\linewidth]{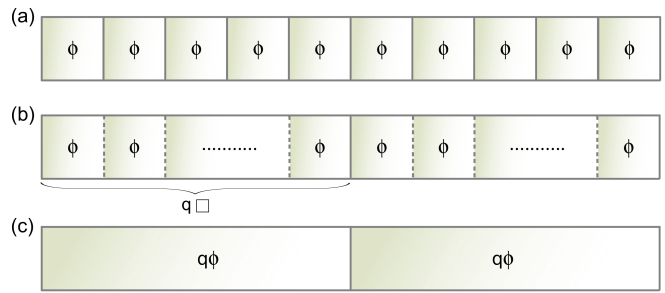}
\end{center}
\vspace{-0.6cm}
\caption{Two-leg ladder lattices: The small circles indicate lattice sites. Solid lines connecting lattice sites indicate bonds along which hopping is allowed. (a) Sketch of 
 uniform two-leg  ladder with flux $\phi$ per elementary  plaquette $\Box$. (b) Sketch of the ladder lattice with explicitly enlarged unit cell of  $q \Box$ plaquettes. 
Along the rungs indicated by dashed lines hopping is blocked. (c) Uniform two-leg  ladder with $q$-times larger lattice constant in the  direction of the legs, 
$q\phi$ flux per plaquette and $q$ times less links along the legs. In particular, if $\phi\in (0,\pi)$ and $q\phi\in (-\pi,0)$, mod $(2\pi)$, the  
chiral current  circulate counterclockwise around the system in (a) and clockwise in (b) and (c).
\label{fig:sketch}}
\end{figure}

As a starting point, consider the uniform two-leg ladder geometry presented in Fig.~\ref{fig:sketch}(a) (i.e., the geometry that corresponds to Eq.~\eqref{eq:ham_rung}) 
and the simplest case of noninteracting bosons. As a modification, in Fig.~\ref{fig:sketch}(b) 
the 
hopping along the rungs indicated by dashed lines is fully suppressed such that the shortest closed path to pick up a phase is the 
boundary of $q$ elementary plaquettes that the system in Fig.~\ref{fig:sketch}(a) is built up from. 
Hence, the relevant flux in this case is $q\phi$. We will argue that the chiral current for the case shown 
in Fig.~\ref{fig:sketch}(b) is expected to be related to the chiral current of a two-leg ladder with $q$ times less plaquetts, but with a flux $q\phi$ per plaquette as shown in Fig.~\ref{fig:sketch}(c). 
If the ladder sketched in Fig.~\ref{fig:sketch}(c) is in the Meissner phase, which is the case for $J_{\bot}>2\tan{\frac{q\phi}{2}}\sin{\frac{q\phi}{2}}$, then its chiral current will be given by
\begin{equation}
\label{MC}
j_c=\frac{J}{q}\sin{\frac{q\phi}{2}},
\end{equation}
where the $1/q$ factor follows from the fact that there are $q$ times less links along the boundary of the ladder 
(i.e., links contributing to the chiral current) for the ladder with  $q$-times less plaquettes as shown in Fig.~\ref{fig:sketch}(c) as compared to the case of  Fig.~\ref{fig:sketch}(a).
As a generalization of the cases shown in Figs.~\ref{fig:sketch}(a) and (b), we introduce a parameter $\delta=\tilde J_{\perp}/J_{\perp}$
where $0\leq \tilde J_{\perp} \leq J_{\perp}$ is the hopping along the dashed links in Figs.~\ref{fig:sketch}(b).
The case $\delta=1$ corresponds to the uniform ladder shown in Fig.~\ref{fig:sketch}(a) and the case $\delta=0$ applies to the case shown in
Fig.~\ref{fig:sketch}(b).
We will next present concrete examples for the minimal cases of $q=2$ and $q=3$.

\subsection{Rung-dimerized ladders: The case of $q=2$}
\begin{figure}[tb]
\begin{center}
\includegraphics[width=1\linewidth]{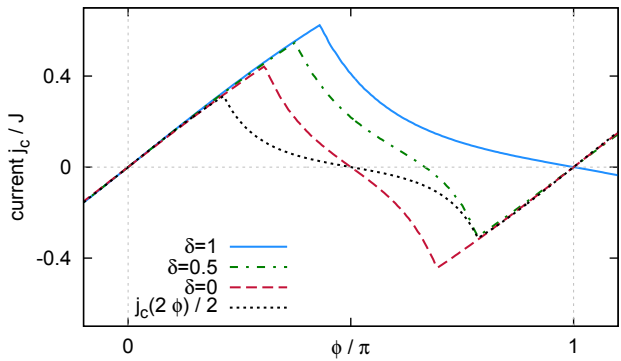}
\end{center}
\vspace{-0.6cm}
\caption{Chiral currents for noninteracting bosons for the cases $\delta=1$,  $\delta=0$, and  flux $q\phi$  of Figs.~\ref{fig:sketch}(a)-(c), respectively,
 for a rung-dimerized ladder $q=2$. We also include results for $\delta=0.5$.
\label{fig:sketch1}}
\end{figure}

First, we consider the minimal case of a rung-dimerized ladder, i.e., $q=2$. In Fig.~\ref{fig:sketch1}, we present the chiral current as a function of flux for $\delta=1,0.5$ and $0$. For comparison, in the same plot we also depict the chiral current corresponding to a uniform ladder with a flux of $2\phi$ per plaquette, but twice less links along the legs corresponding to the situation shown in Fig.~\ref{fig:sketch}(c).

 Obviously, Fig.~\ref{fig:sketch1} shows that the chiral current for $\delta=0$ changes its circulation direction for $0.5<\phi/\pi<1$. 
The agreement between the chiral currents of the cases of Fig.~\ref{fig:sketch}(b) and (c) for $q=2$ is excellent for values of the flux 
 corresponding to the Meissner phase of the model in Fig.~\ref{fig:sketch}(c). Most importantly, the effective uniform ladder with doubled 
flux reproduces correctly the sign of the chiral current for the case of Fig.~\ref{fig:sketch}(b) with $q=2$. 
Clearly, the two curves are not identical because they correspond to two distinct microscopic models.

\subsection{Rung-trimerized ladders: The case of $q=3$}

\begin{figure}[tb]
\begin{center}
\includegraphics[width=1\linewidth]{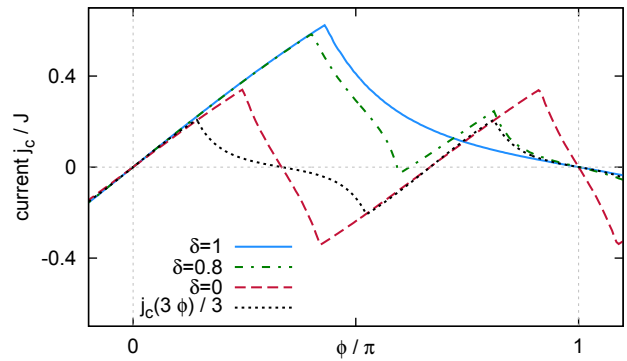}
\end{center}
\vspace{-0.6cm}
\caption{Chiral currents for noninteracting bosons for the cases  $\delta=1$,  $\delta=0$, and flux $q\phi$  of Figs.~\ref{fig:sketch}(a)-(c), respectively, 
 for a rung-trimerized ladder $q=3$. We also include results for $\delta=0.5$.
\label{fig:sketch2}}
\end{figure}

Next, we discuss the rung-trimerized ladder, i.e., $q=3$. In Fig.~\ref{fig:sketch2}, we present the chiral current as a function of flux 
for $\delta=1,0.5,0$. For comparison, in the same plot, we also depict the chiral current corresponding to the 
uniform ladder with a flux of $3\phi$ per plaquette, but a factor of 1/3 less links along the legs as shown in Fig.~\ref{fig:sketch}(c).

The chiral current of noninteracting bosons for $\delta=0$ and for the case of $q=3$ changes its circulation direction for $1/3<\phi/\pi<2/3$. 
Similar to the $q=2$ case, for $q=3$, the agreement between the chiral currents of the cases of Figs.~\ref{fig:sketch}(b) and (c) is excellent for values of the flux 
corresponding to the Meissner phase of the model shown in Fig.~\ref{fig:sketch}(c). Most importantly, the effective uniform ladder with tripled flux reproduces correctly the sign of the chiral current for the case of Fig.~\ref{fig:sketch}(b) with $q=3$.

\subsection{Finite temperatures: Weak-coupling approach}

\begin{figure}[tb]
\includegraphics[width=1\linewidth]{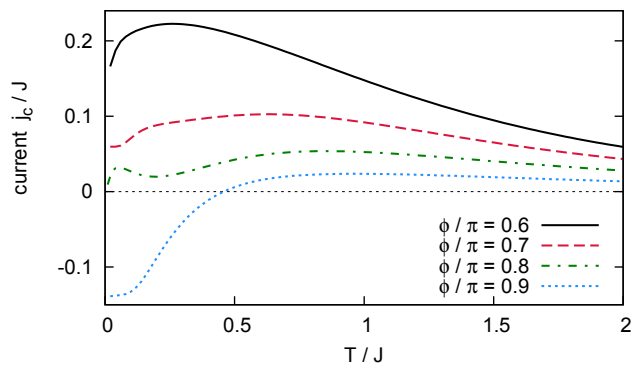}
\caption{Temperature dependence of the chiral current $j_c/J$ within the weak-coupling approximation for various values of the flux $\phi/\pi$.}
        \label{fig:wc_tlarge}
\end{figure}

In order to study the temperature dependence of the chiral current under experimentally realistic conditions \cite{Atala2014}, 
i.e., typically large particle numbers per site, one may use a weak-coupling approximation of model Eq.~\eqref{eq:ham_leg}, 
 introduced in the context of Josephson-junction ladders~\cite{Kardar1986} and applied in the present context in Ref.~\cite{Greschner2015}. 
In the regime of a large filling $\rho\gg 1$ and weak but finite interactions, a suppression of density fluctuations occurs 
and one can thus map the system to a frustrated XY-model of classical spins
\begin{align}
H \to -2J\rho \sum_{\ell =1,2;r=1}^L  \cos( \theta_{\ell,r+1} -\theta_{\ell,r}) \nonumber\\
-2J_{\bot} \rho \sum_{r=1}^L \cos( \theta_{1,r}-\theta_{2,r} -r\phi)\,.
\label{eq:fl_hamiltonianJJL}
\end{align}
This model has been studied using either the effective potentials method~\cite{Griffiths1986,Mazo1995} or a transfer-matrix approach~\cite{Denniston1995} at finite temperatures. We use the latter approach based on the transfer-matrix method~\cite{Denniston1995} directly in the thermodynamic limit to evaluate the chiral current for finite temperatures $T$ ($k_B=1$) through the generalization of the 
Hellman-Feynman-theorem Eq.~\eqref{eq:fl_jc_hellmann}. We compute the derivative of the free energy with respect to the flux
\begin{align}
j_c(\phi)=-\frac{T}{N}\frac{\partial {\ln Z}}{\partial \phi }\,.
\end{align}
Figure~\ref{fig:wc_tlarge} illustrates the temperature dependence of the chiral current for different values of $\phi/\pi$. 
In particular, in the proximity of the VL$_{1/2}$-SF phase (see the $\phi=0.9\pi$ curve of Fig.~\ref{fig:wc_tlarge}), 
it is possible to observe a  chiral-current reversal up to temperatures of the order of $T \approx J/2$ (for more details, see Ref.~\cite{Greschner2015}).
Interestingly, $j_c$ may exhibit a local maximum at finite temperatures, which is related to the frustration of the model and may also be observed for noninteracting particles.

It is important to note that the transfer-matrix technique does not capture the correct high-temperature behavior of the model \eqref{eq:fl_hamiltonianJJL} due to the assumed mapping to a one-dimensional chain. Nonetheless, the leading temperature dependence for $T\gg J$ 
still comes our correctly, with  a decay of the current as $j_c \sim {T}^{-3}$. 
In fact, also for noninteracting particles, one finds a similar decay
\begin{align}
j_c(T) = \frac{ J_\perp^2 \sin(\phi)}{6 (T)^{3}} + \mathcal{O}(T^{-5}) \,.
\end{align}

\subsection{Phenomenological chiral-current curve in vortex-lattice states of weakly interacting bosons}

We now use the method discussed in the previous section to study the high-density regime and very low temperatures.  
We thus consider bosons on a uniform ladder (i.e., the geometry of Fig.~\ref{fig:sketch}(a)) but in the weakly-interacting regime $\rho\gg 1$ and $U\ll J\rho$. 
In \cite{Greschner2015}, we have shown that, for example, for $J_{\perp}=0.5J$, there are pronounced vortex-lattice states (for the smallest temperature 
that we could access in the transfer-matrix approach) for the following vortex densities: $\rho_v=0,1/2,2/5,1/3,1/4,1/5$ (for which 
the unit cell  thus consists of  $q=1,2,5,3,4,5$ plaquettes, correspondingly). Other vortex lattices are washed out already by a small temperature, which we
cannot avoid in the transfer-matrix approach. 
The vortex-lattice state with $q=1$ is the Meissner state.

Now, in the weak-coupling regime, we can reconstruct a chiral-current curve for those flux values, for which vortex-lattice states are realized from just knowing the vortex-density curve as a function of flux  by the following method: in the vortex-lattice states, 
we use the expression  for the chiral current of free bosons with a correspondingly enlarged unit cell. Hence, in the vortex-lattice 
states with $q$ times increased unit cell we will use the  expression  Eq.~(\ref{MC}), 
provided that $J_{\bot}>2J\sin{\frac{q\phi}{2}}\tan{\frac{q\phi}{2}}$ (which happens to be the case for all vortex-lattice states that we observe). 

\begin{figure}[tb]
\begin{center}
\includegraphics[width=1\linewidth]{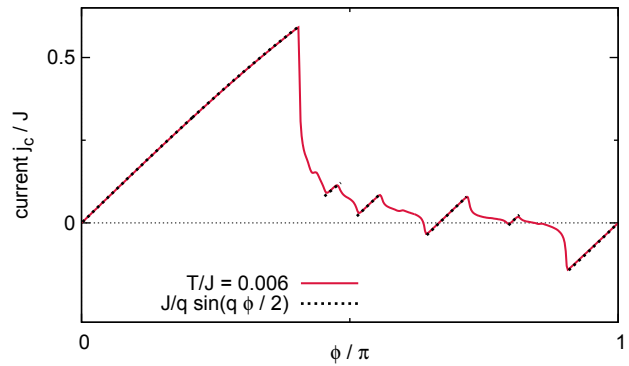}
\end{center}
\vspace{-0.6cm}
\caption{ Comparison of the behavior of the chiral current in vortex-lattice states obtained for weakly-interacting bosons in the high-density limit at small temperature by the transfer matrix approach (continuous curve) to Eq.~(\ref{MC}) (dotted curves).
\label{fig:Comparison}}
\end{figure}

In Fig.~\ref{fig:Comparison}, we see that in vortex-lattice states with a $q$-times enlarged unit cell the behavior of the 
chiral current is captured well by Eq.~(\ref{MC}).
Hence, we conclude that the response of the chiral current of bosons to a spontaneous increase of the unit cell in the weak-coupling regime is not only qualitatively but also quantitatively similar to the response of the chiral current of noninteracting bosons to an explicit enlargement of the unit cell.

\section{CDW phase}
\label{sec:CDW}

At quarter filling $\rho=1/4$ and for sufficiently large $J_\perp,U>J$, a fully gapped CDW$_{1/4}$ phase with a spontaneously broken translational symmetry and a two-fold enlarged unit cell can be observed.
We first reported evidence and a theoretical explanation for this state in \cite{Piraud2015}. 
An example for the typical configuration of currents and density  with staggered rung-density oscillations is sketched in Fig.~\ref{fig:patterns}(f) (the data are compiled from the 
central part of a system with $L=80$ rungs, $\phi=0.98\pi$, $J_\perp=3J$). The ground-state currents look Meissner-like, the rung-currents being suppressed. 

As initially described in Ref.~\cite{Piraud2015}, the emergence of this CDW$_{1/4}$ phase is best understood from the limit of strong interchain tunneling $J_\perp/J \to \infty$. 
By introducing a pseudo-spin-$1/2$ degree of freedom on a rung $r$ via
\begin{align}
  | \uparrow \rangle_r &\mapsto (|1,0 \rangle_r +|0,1 \rangle_r)/\sqrt{2} \nonumber\\
  | \downarrow \rangle_r   &\mapsto |0,0 \rangle_r \,,
\end{align}
one may write down an effective spin-$\frac{1}{2}$ model. To first order in $1/|J_{\bot}|$, the Hamiltonian is $H_{\frac{1}{2}} = J\, H_{\frac{1}{2}}^0 + J^2/|J_{\bot}|\, H_{\frac{1}{2}}^1$:
with 
\begin{align}
\label{eq:fl_effectivespinhalf}
H_{\frac{1}{2}}^0 &= \cos\left(\frac{\phi}{2}\right)\, \sum_r S^+_r S^-_{r+1} + h.c.\nonumber\\
H_{\frac{1}{2}}^1 &= -\cos\left(\frac{\phi}{2}\right)^2\, \sum_r S^+_r (1/2 + S^z_{r+1}) S^-_{r+2} + h.c.\nonumber\\
&\quad-\frac{1}{2}\sin\left(\frac{\phi}{2}\right)^2\, \sum_r S^+_r (1/2 - S^z_{r+1}) S^-_{r+2} + h.c. \nonumber\\
&\quad-\frac{1+3 \cos\left(\phi\right)}{2}\, \sum_r S^z_r S^z_{r+1}\,.
\end{align}
While for small fluxes and in this effective model, the term $H_{\frac{1}{2}}^0$ dominates and describes a usual ($c=1$) Luttinger-liquid phase,
corresponding to the M-SF phase, for fluxes $\phi\to\pi$, the correlated hopping and nearest-neighbor Ising-type interaction terms become relevant. At $\phi=\pi$,
 we may simplify the effective model to
\begin{align}
H_{\frac{1}{2}} = - \frac{J^2}{2|J_{\bot}|} \sum_r \left\lbrack S^+_r \left(\frac{1}{2} - S^z_{r+1}\right) S^-_{r+2} + {\rm h.c.} \right. \nonumber \\
		\left. - 2 S^z_r S^z_{r+1}\right \rbrack \,.
\label{eq:fl_effH12pi}
\end{align}
Since for a large filling and due to the correlated hopping basically all tunneling processes are strongly suppressed, 
the Ising term $S^z_r S^z_{r+1}$ induces a transition to a doubly degenerate N\'eel state at quarter filling ($\rho=1/4$)  and in the vicinity of $\phi=\pi$. 
For the original bosonic particles, this corresponds to a CDW$_{1/4}$ phase.

\begin{figure}[tb]
\includegraphics[width=1\linewidth]{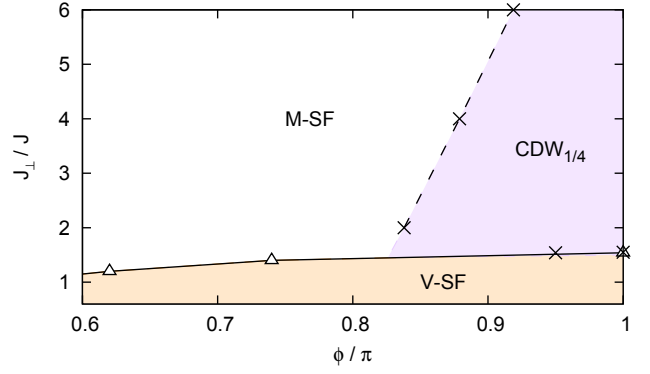}
\caption{Phase diagram for $\rho=0.25$ and hard-core bosons ($U/J\to\infty$) as a function of $\phi/\pi$ versus $J_\perp/J$. 
Symbols represent estimated points of the Meissner-to-vortex ($\Delta$) and the Meissner-to-CDW$_{1/4}$ ($\times$) phase transitions (see the discussion in Sec.~\ref{sec:CDW}).
Straight lines and shadings are a guide to the eye.}
\label{fig:pd_HBC_n0.25}
\end{figure}

In Fig.~\ref{fig:pd_HBC_n0.25}, we show the phase diagram of the CDW$_{1/4}$ phase as a function of the flux $\phi/\pi$ and $J_\perp/J$ in the limit of hard-core bosons $U/J\to\infty$. 
We estimate the position of the SF-to-CDW$_{1/4}$ transition by calculating the Luttinger-liquid parameter $K_\rho$, which at the transition should be $K_\rho=1/2$. 
The CDW$_{1/4}$ phase remains stable for fluxes $\phi\gtrsim 0.8\pi$ and $J_\perp\gtrsim 1.5J$.

\begin{figure}[tb]
\includegraphics[width=1\linewidth]{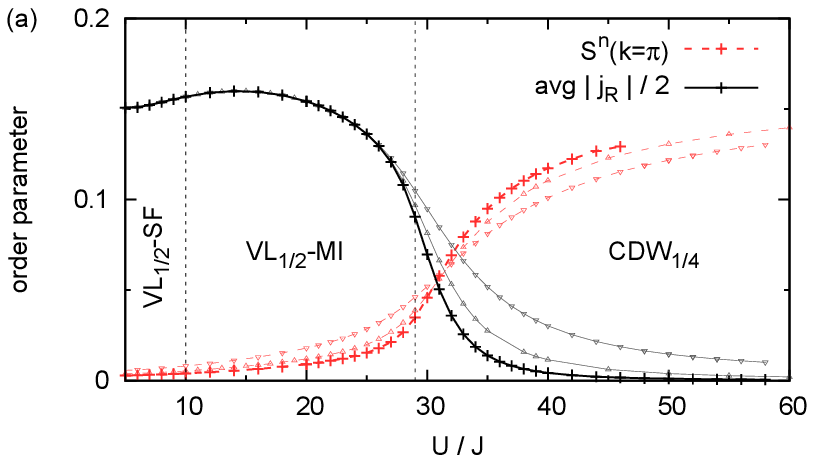}
\includegraphics[width=1\linewidth]{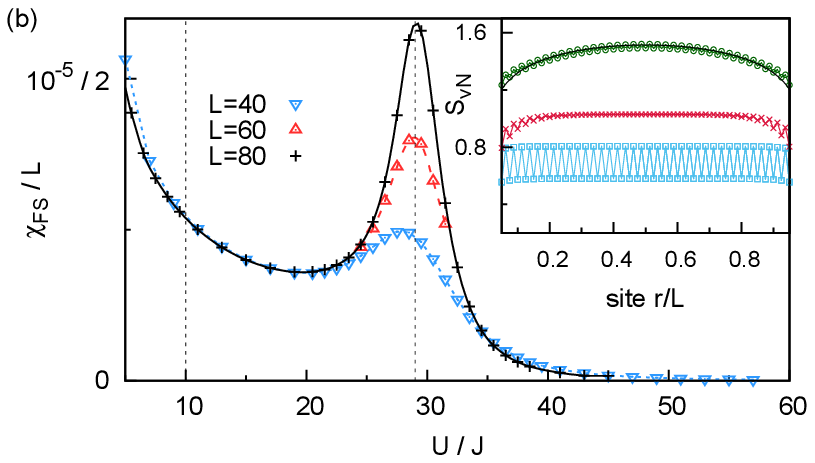}
\caption{CDW$_{1/4}$ phase: Emergence of insulating phases at quarter filling $\rho=0.25$ as a 
function of $U/J$ for $\phi=0.98\pi$ and $J_\perp=3.2 J$.
(a) Average rung current ${\rm avg} | j_R |$ and CDW$_{1/4}$ order parameter $S^n(k=\pi)$, i.e., the value of the static density structure factor $S^n(k=\pi)$, 
for several system sizes $L=80$ ($+$ symbols), $L=60$ ($\Delta$) and $L=40$ ($\nabla$).
(b) Scaling of the fidelity susceptibility $\chi_{FS}/L$ for $L=40$, $60$ and $80$ rungs. The inset shows examples for the entanglement entropy $S_{\rm vN}(r)$ for 
(from top to bottom) the VL$_{1/2}$-SF ($U=J$, with a  fit of Eq.~\eqref{eq:ee_obc} for $c=1$ to the data indicated by the black solid line), 
CDW$_{1/4}$ ($U=58J$) and VL$_{1/2}$-MI ($U=20J$).
Dashed lines indicate the estimated locus of phase transitions.
}
\label{fig:pd_Ucut_n0.25}
\end{figure}

We next analyze the stability of the CDW$_{1/4}$ phase at finite interactions $U/J <\infty$ (see Fig.~\ref{fig:pd_Ucut_n0.25}). For small $U/J$, 
we expect a VL$_{1/2}$-SF phase for sufficiently large fluxes $\phi\to\pi$. As shown in Fig.~\ref{fig:pd_Ucut_n0.25}, remarkably, we observe a large regime of a fully 
gapped VL$_{1/2}$-MI phase with finite staggered rungs currents and a flat entanglement entropy profile $S_{\rm vN}(r)$ (see the inset of Fig.~\ref{fig:pd_Ucut_n0.25}(b)). 
In this MI, one particle is delocalized in each plaquette, with suppressed charge fluctuations between plaquettes. This state results
from applying the flux and is not linked to a trivial band insulator in the absence of the flux. 

For $J_\perp=3.2J$, we estimate the BKT-transition point from the VL$_{1/2}$-SF to the VL$_{1/2}$-MI phase to be at $U\approx 10 J$,  again from 
determining the point at which the Luttinger-liquid parameter becomes $K_\rho=1$.  
For $U\gtrsim 30 J$, we finally observe the CDW$_{1/4}$ phase, with vanishing ${\rm avg} |j_R|$ and a finite staggered charge-density-wave order, indicated by the peak-value of 
the static density structure factor $S^n(k=\pi)$ (see Fig.~\ref{fig:pd_Ucut_n0.25}(a)).
The possibly Gaussian phase transition can be located precisely from the pronounced peak of the fidelity susceptibility $\chi_{FS}/L$,
 which we find to diverge as ${\rm max}\left(\chi_{FS}/L\right) \sim L^{3/2}$.

We also study the CDW$_{1/4}$ phase for systems with a synthetic lattice dimension  including the rung interactions Eq.~(\ref{eq:ham_Vrung}) for the case $U=V$. It is important to note
 that we observe broad regimes of both stable CDW$_{1/4}$ and VL$_{1/2}$-MI phases for similar parameter ranges as in Fig.~\ref{fig:pd_Ucut_n0.25} for the $V=0$ case (data for $V>0$ 
not shown here). Our preliminary results, however, suggest the possibility of an intermediate V-MI phase for $U=V\sim 30J$ for $V>0$.  
These results will be published elsewhere.

\begin{figure}[tb]
\includegraphics[width=1\linewidth]{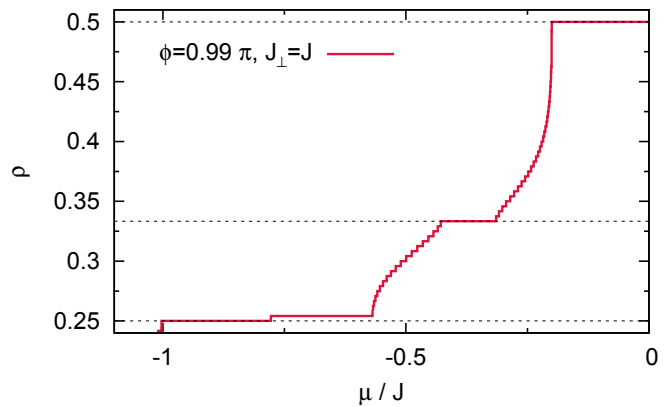}
\caption{Equation of state $\rho=\rho(\mu)$ for $U/J\to\infty$, $V/J\to\infty$ and $J_\perp=J$, $\phi=0.99\pi$ ($L=120$ rungs). At fillings $\rho=1/4$, $\rho=1/3$ and $\rho=1/2$ (indicated by dashed lines) extended plateaus correspond to the gapped CDW$_{1/4}$, CDW$_{1/3}$ and MI phase, respectively.}
\label{fig:RCB_murho_tp1.0_phi0.99}
\end{figure}

In Fig.~\ref{fig:RCB_murho_tp1.0_phi0.99}, we present the equation of state $\rho=\rho(\mu)$ for the case of strong nearest-neighbor rung and onsite interactions, $V/J\to \infty$ and $U/J\to\infty$. Hence, in this limit, only a maximal occupation of a single particle per rung is allowed. The DMRG simulation shows that, due to the strong interactions, already for small interchain tunneling $J_\perp \sim J$ an extended CDW$_{1/4}$ phase is stabilized at quarter filling $\rho=1/4$. Very interestingly, close to the limit of $\phi=\pi$ (see  Fig.~\ref{fig:RCB_murho_tp1.0_phi0.99}),  a CDW$_{1/3}$ also emerges at filling $\rho=1/3$. This state has a three-fold enlarged unit cell and a density oscillation corresponding to a N\'eel-state of type $|\cdots \uparrow\uparrow\downarrow \uparrow\uparrow\downarrow \cdots\rangle$ in the rung-singlet basis. A detailed analysis of these CDW-states in the rung-hard-core limit will be published elsewhere.

\section{Biased-ladder phase}
\label{sec:BLP}

In this section, we turn to the discussion of the biased-ladder phase (BLP), first discussed using mean-field
theory by Wei and Mueller \cite{Wei2014}.
In the BLP phase, the $Z_2$ symmetry associated with interchanging the leg index $\ell =1 \to 2$ and $\ell = 2\to 1$
(or, in other words, reflection symmetry with respect to reflections about the middle of each rung) is spontaneously
broken. This results in a density imbalance $\Delta n$ between the two legs, which serves as the order parameter
for this phase. We define $\Delta n $ as 
\begin{align}
\Delta n = \sum_r |\left< n_{1,r}-n_{2,r}\right>| / N \;.
\label{eq:blpdeltan}
\end{align}
The BLP phase was studied previously in several works \cite{Wei2014,Uchino2015,Uchino2016,Greschner2015}. In  \cite{Greschner2015},
we established the existence of the BLP phase at intermediate values of $U/J\sim 2$ from DMRG simulations for $\rho \lesssim 1$.
Here, we will apply a theoretical framework valid in the regime of a dilute Bose gas to describe this state and to obtain the 
phase diagram in the dilute-gas limit. That theory relies on a mapping of the system to a two-component Lieb-Liniger gas, whose
parameters we relate to microscopic parameters by studying the scattering problem. 
This theory is described in Sec.~\ref{sec:blp_dilute}, while we complement the analytical analysis by DMRG results for finite densities
$\rho \lesssim 1$ presented in Sec.~\ref{sec:blp_dmrg}.
 
\subsection{Biased-ladder phase in the dilute limit}
\label{sec:blp_dilute}

In this section, we  address the limit of a dilute gas of bosons,
which in one dimension is a strong-coupling regime, invalidating a mean-field type approach. 
Moreover, it is not possible to develop an effective field-theory approach based on bosonization 
because the velocities obtained from linearizing the single-particle dispersion vanish together with the density. 
We will follow an approach that we developed for frustrated one-dimensional spin systems close to their saturation magnetization \cite{Kolezhuk2012,Arlego2011}, a method that works qualitatively the best in the dilute limit. 

For convenience, we use the leg gauge  to render the 
system explicitly translationally invariant. Thus (quasi-)momentum is a good quantum number. 
In that gauge, the Hamiltonian is given by $H_{\rm leg}$ from Eq.~\eqref{eq:ham_leg}.

\subsubsection{Single-particle dispersion}

First, we study the single-particle dispersion on a ladder with a nonzero flux. 
The dispersion consists of two branches (or bands), labelled by $\pm$: 
\begin{eqnarray}
\epsilon_{\pm}(k)= -2\cos{k}\cos{\frac{\phi}{ 2}}\pm \sqrt{   J^2_{\bot} +4\sin^2\frac{\phi}{2}\sin^2k }.
\end{eqnarray}

\begin{figure}[tb]
\includegraphics[width=1\linewidth]{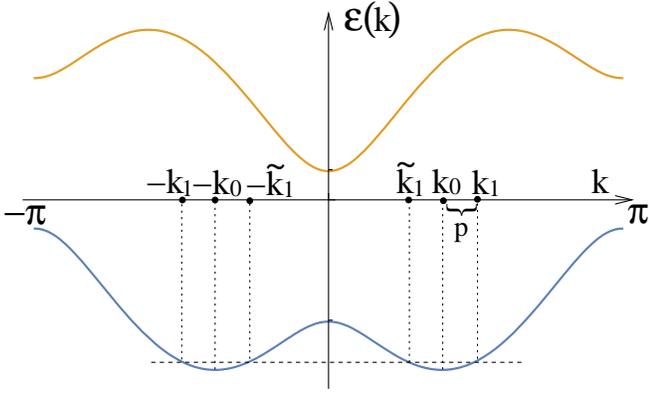}
\caption{ Two branches of the single-particle dispersion with doubly degenerate minima at $\pm k_0$ 
in the lowest band for parameters $J_{\bot}=J$ and $\phi=0.8\pi$. $k_0,k_1$ and $\tilde k_0,\tilde k_1$ are the incoming and outgoing momenta in the two-body scattering problem described in Sec.~\ref{sec:scatt}.}
\label{fig:disp}
\end{figure}

For small values of fluxes, $J_{\bot}>2\tan{\frac{\phi}{2}}\sin{\frac{\phi}{2}}$,
the  lower single-particle band has a unique minimum at $k=0$, corresponding to the Meissner phase \cite{Orignac2001}. 
It develops a double minimum for $J_{\bot}<2\tan{\frac{\phi}{2}}\sin{\frac{\phi}{2}} $ at two points $\pm k_0$ as depicted in Fig.~\ref{fig:disp},
\begin{equation}
\pm k_{0}=\pm \arcsin \sqrt{ \sin^2({{\phi}/{2}})  -\frac{J^2_{\bot}}{4\tan^2({{\phi}/{2}})}  }.
\end{equation}
In the  free case $U=0$, the many-body ground state is infinitely degenerate for periodic boundary conditions. Interactions can lift this degeneracy,
which leads to the many interesting quantum phases detected in this model.
 The relevant question is whether the interacting system prefers to form
a condensate that is a fifty-fifty mixture of condensates at each of the minima or to exclusively populate  one of the two equivalent minima in the 
dispersion.

\subsubsection{Two-component Lieb-Liniger model}

We  next follow an approach that we have developed for the dilute Bose-gas limit \cite{Arlego2011}, using  a mapping of the original  bosonic system with a 
doubly-degenerate single-particle dispersion to a two-component Lieb-Liniger model
\begin{eqnarray}
\label{LiebLiniger}
H_{\rm eff}=\int &\mathrm{d}x & \left[-\hat \Psi^{\dagger}_1(x) \frac{\partial_x^2}{2m}\hat \Psi_{1}(x)-\hat \Psi^{\dagger}_2(x) \frac{\partial^{2}_x}{2m}\hat\Psi_{2}(x)
\right. \nonumber \\&& \left. +\frac{g}{2}(n^{2}_1+n^{2}_2) 
+ \tilde gn_1n_2\right]
\end{eqnarray}
where $\Psi_{1,2}(x)$ are boson field operators corresponding to particles from left and right dispersion minima
and $n_{1,2}$ are the corresponding density operators. 
The coupling constants of the effective two-component Lieb-Liniger model are given by
\begin{equation}
\label{CCSL}
g=-\frac{2}{ma}\,\,\mathrm{ and}\,\, \tilde g=-\frac{2}{m\tilde a},
\end{equation}
 where $a$ and $\tilde a$ are intra- and inter-species scattering lengths, respectively, related to the 
corresponding scattering phase shifts via
\begin{equation}
\label{SL}
 a=\lim_{p\to 0}\frac{\cot{\delta}}{p}\,\,  \mathrm{and} \,\, \tilde a=\lim_{p\to 0}\frac{\cot{\tilde \delta}}{p}. 
\end{equation}
In Eq.~(\ref{SL}), $p$ is the relative momentum of the low-energy two-boson scattering problem modulo $k_0$. The 
effective mass $m$ is the same for both species
\begin{equation}
\label{mass}
m=\frac{\sin{\frac{\phi}{2}}\tan^2{\frac{\phi}{2}}\sqrt{J_{\bot}^2+4\sin^2{\frac{\phi}{2}}} }  { 4\sin^2{\frac{\phi}{2}}   \tan^2{\frac{\phi}{2}} -J_{\bot}^2}
\end{equation}
and it diverges at the Lifshitz transition when the two minima of the single-particle dispersion at $\pm k_0$ merge into a single one at $k=0$.

From the effective model Eq.~(\ref{LiebLiniger}) it follows that for $g<\tilde g$, the ground state corresponds to a 
single-component Luttinger-liquid state (thus an immiscible state where only one species of  bosons are present) with a spontaneously broken $Z_2$ symmetry, 
whereas for $g>\tilde g$, energetically a two-component Luttinger-liquid state (and thus a miscible state with both species of bosons present in the ground state) 
is preferred. The phase transition line between these two states is given by $g=\tilde g$ and the phase transition is first order.

\subsubsection{Scattering problem}
\label{sec:scatt}

We will extract the relevant scattering lengths from solving the low-energy two-boson scattering problem on-shell, hence neglecting the upper dispersion branch in Fig.~\ref{fig:disp}. The scattering state of two particles ($i=1,2$) with momenta $k_1$ and $k_2$ has the following energy
\begin{equation}
E=\sum_{i=1,2}\left(-2\cos{k_i}\cos{\frac{\phi}{ 2}}-\sqrt{   J^2_{\bot} +4\sin^2\frac{\phi}{2}\sin^2k_i } \right)\,.
\end{equation}
The two particle wave-function is represented as
\begin{eqnarray}
| \psi \rangle &=&\sum_{i\le j}( C^{1,1}_{i,j}a^{\dagger}_{1,i}a^{\dagger}_{1,j}|0 \rangle+C^{2,2}_{i,j}a^{\dagger}_{2,i}a^{\dagger}_{2,j}|0 \rangle) \nonumber\\
&+&\sum_{i,j}C_{i,j} a^{\dagger}_{1,i}a^{\dagger}_{2,j}|0 \rangle .
\end{eqnarray}
The upper indices on the amplitudes $C^{\ell,\ell}_{i,j}$ indicate that both particles belong to the same leg $\ell$. If there are no such upper indices, then those amplitudes correspond to the case in which 
one particle is on the first leg and the other one on the second leg.

Introducing the total momentum $\Lambda=k_1+k_2$, we separate the center-of-mass motion
\begin{equation}
C^{\ell,\ell}_{i,j}=e^{i\frac{i+j}{2}\Lambda}C^{\ell,\ell}_r, \quad C_{i,j}=e^{i\frac{i+j}{2}\Lambda}C_r,
\end{equation}
where we introduced a relative coordinate $r=j-i$.
The Schr\"odinger equation
\begin{equation}
H | \psi (\Lambda) \rangle =E  | \psi (\Lambda)\rangle
\end{equation}
leads to the following system of equations for the amplitudes $C^{\ell,\ell}_r$, $C_{ r}$, and $C_{ -r}$ for $r>1$,
\begin{eqnarray}
\label{ampl1}
EC^{\ell,\ell}_r=&-&2\cos{(\frac{\phi}{2} +\! (-1)^{\ell}\frac{\Lambda}{2} )}( C^{\ell,\ell}_{r-1}\!+C^{\ell,\ell}_{r+1} )\nonumber\\
&-&J_{\bot}( C_r+C_{-r} ) \nonumber\\
EC_r= &-&2\cos{\frac{\Lambda}{2}}[ e^{i\frac{\phi}{2}}C_{r-1}\!+\! e^{-i\frac{\phi}{2}}C_{r+1}   ] \nonumber\\
     &-&J_{\bot}( C^{1,1}_r\!+C^{2,2}_r) \nonumber\\
EC_{-r}=&-&2\cos{\frac{\Lambda}{2}}[ e^{i\frac{\phi}{2}}C_{-r-1}\!+ e^{-i\frac{\phi}{2}}C_{-r+1}   ]\nonumber\\
&-&J_{\bot}\!( C^{1,1}_r\!+C^{2,2}_r)
\end{eqnarray}
and to the following system of equations for $r \le 1$,
\begin{eqnarray}
\label{ampl2}
(E-U)C^{\ell,\ell}_0=&-&2\cos{(\frac{\phi}{2} + (-1)^{\ell}\frac{\Lambda}{2} )} C^{\ell,\ell}_{1}\nonumber\\
&-&J_{\bot}C_0 \nonumber\\
EC^{\ell,\ell}_1=&-&2\cos{(\frac{\phi}{2} + \!(-1)^{\ell}\frac{\Lambda}{2} )}( 2C^{\ell,\ell}_{0}+C^{\ell,\ell}_{2} )\nonumber\\
&-&J_{\bot}( C_1+C_{-1} ) \nonumber\\
(E-V)C_0=&-&\!2 \cos{\frac{\Lambda}{2}}(e^{-i\frac{\phi}{2}}C_1+\! e^{i\frac{\phi}{2}}\!C_{-1})\nonumber\\
&-&2J_{\bot}(C^{1,1}_0\!\!+\!C^{2,2}_0 )\nonumber\\
EC_{\pm 1}= &-&2 \cos{\frac{\Lambda}{2}}( e^{\pm i\frac{\phi}{2}}C_0 + e^{\mp i\frac{\phi}{2}}C_{\pm 2} ) \nonumber\\
&-&J_{\bot}(C^{1,1}_1+C^{2,2}_1 ).
\end{eqnarray}
From the structure of Eqs.~(\ref{ampl1}) and Eqs.~(\ref{ampl2}), 
it follows that $C^{\ell,\ell}_r$ are real for all $r\ge 0$, whereas $C_{-r}=C^*_r$ and, in particular, $C_0$ is real.

To extract the intra-species scattering length we set $k_1=k_0+p_1$ and $k_2=k_0+p_2$ and take the limits $p_1\to 0$ and $p_2\to 0$. 
Next, we define the relative momentum $p=(k_1-k_2)/2$ and construct scattering states for $r\ge 1$ as follows
\begin{eqnarray}
\label{ansatz}
C_r^{1,1}\!\!&=&\!2\cos{\theta_{k_1}}\cos{\theta_{k_2}}\cos{(pr+\delta)}+v_1(a_1z_1^r+  a^*_1(z_1^*)^r)   \nonumber   \\
&+&v_2(a_2z_2^r+  a^*_2(z_2^*)^r)+v_3(a_3z_3^r+  a^*_3(z_3^*)^r)\nonumber\\
C_r^{2,2}&=&2\sin{\theta_{k_1}}\sin{\theta_{k_2}}\cos{(pr+\delta)}+v_1(b_1z_1^r+  b^*_1(z_1^*)^r)\nonumber\\
&+&v_2(b_2z_2^r+  b^*_2(z_2^*)^r)+v_3(b_3z_3^r+  b^*_3(z_3^*)^r)\nonumber\\
C_r&=&\cos{\theta_{k_1}}\sin{\theta_{k_2}}e^{i(pr+\delta)}+ \sin{\theta_{k_1}}\cos{\theta_{k_2}}e^{-i(pr+\delta)} \nonumber\\
&+& \! v_1(c_1z_1^r+  (z_1^*)^r) +\!v_2(c_2z_2^r+ (z_2^*)^r) \nonumber\\
&+&v_3(c_3z_3^r+ \! (z_3^*)^r),
\end{eqnarray}
where $\delta$ is the scattering phase shift. Furthermore, we introduced Bogoliubov coefficients in analogy to the free boson case
\begin{equation}
\theta_{k_i}= \frac{1}{2} \arctan\left[ \frac{J_{\bot}}{\cos{(k_i-\phi/2) }   -\cos{(k_i+\phi/2)}} \right].
\end{equation}
The scattering states of Eq.~(\ref{ansatz}) should, for large relative distance $r\gg 1$ and for $\delta=0$, 
 reproduce the scattering states of noninteracting bosons $U=0$, hence $|z_i|<1$ for $i=1,2,3$.
The real numbers $v_1,v_2$ and $v_3$ will be fixed later together with the scattering phase shift $\delta$. 
First, we insert the ansatz given by Eq.~(\ref{ansatz}) into the 
Eqs.~(\ref{ampl1}) to determine the complex coefficients $a,b,c$ and $z$, where, for physically acceptable solutions, we require $|z_i|<1$.
This leads to the following set of equations:
\begin{eqnarray}
Ea+J_{\bot}(c+1)+2a\cos{(\frac{\phi}{2}-\frac{\Lambda}{2})(z+1/z)=0}\nonumber\\
Eb+J_{\bot}(c+1)+2b\cos{(\frac{\phi}{2}+\frac{\Lambda}{2})(z+1/z)=0}\nonumber\\
E+2\cos{\frac{\Lambda}{2}} ( e^{-i\frac{\phi}{2}}z+e^{i\frac{\phi}{2}}/z )  +J_{\bot}(a+b)/c=0\nonumber\\
E+2\cos{\frac{\Lambda}{2}} ( e^{i\frac{\phi}{2}}z+e^{-i\frac{\phi}{2}}/z )  +J_{\bot}(a+b)=0\,.
\end{eqnarray}
Taking only physically meaningful solutions of this system of equations (in general, there are exactly three such solutions) 
for $a_i,b_i,c_i$ and $z_i$ for $i=1,2,3$,  we insert the  ansatz Eq.~(\ref{ansatz}) into Eq.~(\ref{ampl2}) to solve for the unknown quantities 
$\delta, C^{1,1}_0, C^{2,2}_0, C_0, v_1,v_2,$ and $v_3$.

Along similar lines, we extract the scattering phase shift for the inter-species scattering $\tilde \delta$. In that case, 
we can fix total momentum to zero $\Lambda=0$ by considering one boson with momentum $k_1=k_0+p$, $p\to 0$, and another one 
with $-k_1$. Hence, one can search for solutions satisfying $C^{1,1}_r=C^{2,2}_r$, reducing the number of unknown constants. 
However, one needs to take into account the fact that low-energy scattering states for inter-species scattering with a given 
total momentum and energy are doubly degenerate. There is no such degeneracy for the intra-species scattering problem outlined above: 
for intra-species scattering the scattering state is uniquely characterized by total momentum and energy. 
From the form of the single-particle dispersion,  the  origin of the degeneracy of the inter-species scattering problem is obvious. 
For example, the scattering state of zero total momentum and asymptotic momenta $k_1$ and $-k_1$ is clearly degenerate with the scattering state 
of two bosons with momenta $\tilde k_1$ and $-\tilde k_1$ as indicated in Fig.~\ref{fig:disp}. When constructing the scattering state 
for inter-species scattering one therefore has  to admix different asymptotic momenta: the incoming state with momenta $k_1$ and $-k_1$ will produce a state with similar outgoing momenta superposed with a scattering state with outgoing momenta $\tilde k_1$ and $-\tilde k_1$.
Denoting $k_2=-k_1$ and $\tilde k_2=-\tilde k_1$,  we write the following ansatz for inter-species scattering for $r\ge 1$,
\begin{eqnarray}
\label{interansatz}
&&C_r^{l,l}=-2\cos{\Theta_{k_1}} \cos{\Theta_{k_2}}\cos(k_1 r+\tilde \delta)+ vz^r\nonumber \\
&&-2v_1 \cos{\Theta_{\tilde k_1}} \cos{\Theta_{\tilde k_2}}\cos(\tilde k_1 r-\tilde \delta)\nonumber\\
&&C_r\!=\!\cos{\Theta_{k_1}} \sin{\Theta_{k_2}}e^{-i(k_1 r+\tilde \delta)} \!\!+ \sin{\Theta_{k_1}} \!\cos{\Theta_{k_2}}e^{i(k_1 r+\tilde \delta)}\nonumber\\
&&+ v_1\!\left( \cos{\Theta_{\tilde k_1}}\! \sin{\Theta_{\tilde k_2}}\!e^{-i(\tilde k_1 r-\tilde \delta)}\!+\! \sin{\Theta_{\tilde k_1}}\! \! \cos{\Theta_{\tilde k_2}}\!e^{i(\tilde k_1 r-\tilde \delta)} \!  \right)\nonumber\\
&&+vv_2z^r \nonumber\\
&&C_{-r}=C_r^*,
\end{eqnarray}
where $v, v_1,$ and $z$ are real numbers with $|z|<1$ to ensure  physically meaningful solutions and $v_2$ can be any complex number. 
We note that since $k_1$ and $k_2$ (as well as $\tilde k_1$ and $\tilde k_2$) 
have opposite sign one has to take the proper branches of the $\arctan$ in the definition of the coefficients $\Theta_{k_i}$.
The unknowns $v_2$ and $z$ are fixed by inserting Eq.~(\ref{interansatz}) into the system of Eqs.~(\ref{ampl1}).
In particular, for $z$, we obtain
\begin{equation}
z=-\frac{E+\sqrt{E^2-16\cos^2{(\phi/2)}  }}{4\sin{(\phi/2)}}
\end{equation}
and for $v_2$,
\begin{equation}
v_2=\frac{2iJ_{\bot}\cot{(\phi/2)}}{\sqrt{E^2-16\cos^{2}{(\phi/2)}}}.
\end{equation}

The remaining five unknowns  $\tilde \delta, C^{1,1}_0= C^{2,2}_0, C_0, v,$ and $v_1$ are determined by inserting the 
ansatz Eq.~(\ref{interansatz}) into the system of Eqs.~(\ref{ampl2}). 

With the help of relations Eq.~(\ref{CCSL}) and Eq.~(\ref{SL}) we determine the intra- and inter-species scattering lengths and the 
corresponding Lieb-Liniger coupling strengths.
For $g>\tilde g$, both minima at $\pm k_0$ are equally populated in the ground state  and a 
two-component Luttinger-liquid phase is realized for a finite but small density. This is the vortex-superfluid phase. 
There is no density imbalance between the legs of the ladder  since $C^{1,1}_r=C^{2,2}_r$ for the inter-species scattering problem. 
However, for $g<\tilde g$, only one of the minima in the single-particle dispersion  is populated and a one-component phase is selected, where a density imbalance between the two legs exists, which, for the $U\to 0$ limit is given by,
\begin{equation}
\frac{\delta \rho}{ \rho}=\cos{2\Theta_{k_0}}.
\end{equation}
This state is the BLP superfluid.

\subsubsection{Phase diagram with the BLP state in the dilute Bose-gas  regime}

\begin{figure}[tb]
\includegraphics[width=1\columnwidth]{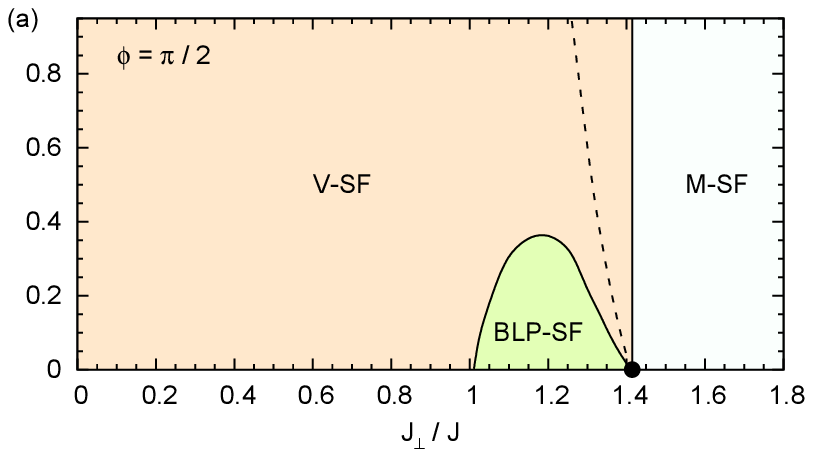}
\includegraphics[width=1\columnwidth]{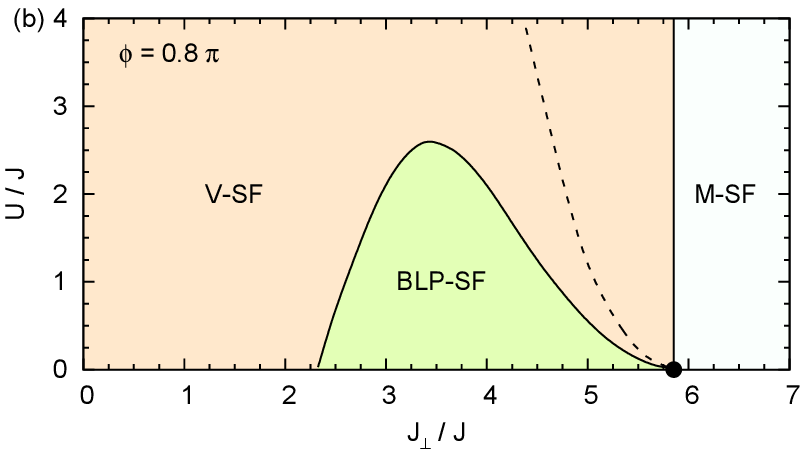}
\caption{ Ground-state phase diagram of the bosonic ladder in the dilute limit for $V=0$ for (a) $\phi=0.5\pi$ and (b) $\phi=0.8\pi$. 
The dashed line does not represent a phase transition, rather it indicates the line where the intra-species scattering length vanishes. Above the dashed line, in the V-SF phase, the intra-species interaction enters into the super-Tonks regime. The Lifshitz point, beyond which the M-SF sets in, 
is indicated by a filled circle at $J_{\bot}=2J\sin{\frac{\phi}{2}}\tan{\frac{\phi}{2}}$. 
}
\label{fig:pdd}
\end{figure}

In Figs.~\ref{fig:pdd}(a) and (b), we present the ground-state phase diagram as a function of $J_{\bot}$ and $U$ obtained in the dilute-gas 
limit for the case of $V=0$ and  for two values of the flux  $\phi=0.5\pi$  and $\phi=0.8\pi$, respectively. 
One can see that the region  in parameter space, in which the BLP-SF phase exists,  
grows significantly 
in the parameter plane $U/J$ versus $J_{\bot}/J$ when increasing the flux from $\phi=\pi/2$ to $\phi=0.8\pi$. 

Apart from the phase transition lines we also indicate the  line (dashed line) above which, in the V-SF phase, 
the intra-species interaction enters into the so-called super-Tonks regime~\cite{Astrakharchik2004, Batchelor2005, Astrakharchik2005} with $a>0$, meaning that intra-species repulsion (i.e., repulsion between the particles with almost the same momenta) is effectively stronger than the hard-core contact repulsion. 
There, instead of relating the scattering length $a$ to the  Lieb-Liniger coupling constant $g$ through Eq.~(\ref{CCSL}) (which 
would wrongly imply attractive $g$), the  intra-species scattering length should be interpreted as an excluded volume.
Inter-species interactions, on the contrary, never enter the super-Tonks regime. The interesting property of the  gauge field is that 
the super-Tonks regime (for intra-species interactions) is attained for the case of contact interactions for finite values of the repulsive interaction. The 
critical value of the repulsive interactions for attaining the super-Tonks regime in intra-species scattering goes to zero when approaching the Lifshitz point,
where the effective mass of the Lieb-Liniger model given in Eq.~(\ref{mass}) diverges.

Including repulsive interactions along the rung with $V>0$ increases the region of stability of the BLP phase. For example, 
for $\phi=0.5 \pi$ and at $J_{\bot}=1.2J$, the transition from the BLP-SF to the V-SF for $V=0$ is at $U\simeq 0.37J$, whereas for $V=U$, 
that transition shifts to $U/J\simeq 0.5$. For $\phi=0.8 \pi$ and at $J_{\bot}=3.5J$, the  transition from the BLP-SF to the V-SF is, 
for $V=0$, at $U/J\simeq 2.65$, whereas for $V=U$, the transition shifts to $U/J\simeq 7$. 

Attractive interactions along the rung $V<0$ with $|V| \ll U$, on the contrary, shrink the region of stability of the BLP-SF, consistent 
with expectations on physical grounds. For stronger attractions, bound states can develop and a pair superfluid can get stabilized with no density imbalance between the legs of the ladder. For example, 
for $\phi=0.8 \pi$, $J_{\bot}=3.5J$ and $U=3J$ there is a resonance in inter-species scattering $\tilde g=0$ at $V=V_c \simeq -1.84J$ and for $V<V_c$,
 instead of the  two-component V-SF, a single-component pair-superfluid phase (P-SF) is stabilized. Further decreasing $V$, the  system 
can eventually collapse. We estimate the  instability to a collapse  to occur  when the intra-species interaction constants become attractive $g<0$.
 For $\phi=0.8 \pi$, $J_{\bot}=3.5J$ and $U=3J$ this happens at $V/J\simeq -2.6$.

\subsection{BLP at finite densities}
\label{sec:blp_dmrg}

We next study the BLP phase using DMRG calculations at finite densities. The current configuration of the BLP phase is very similar to the Meissner phase, as can be seen 
in Fig.~\ref{fig:patterns}(e): the current flows only along the boundary of the ladder while the rung currents are suppressed. 
The particle density, however, exhibits a marked imbalance between the legs, which we  calculate from Eq.~\eqref{eq:blpdeltan}.
 
In the thermodynamic limit the ground state is thus two-fold degenerate, spontaneously breaking the $Z_2$ mirror symmetry between the legs.
In order to numerically stabilize the simulation of the BLP phase, we add small potentials at the boundary of the ladder explicitly breaking the symmetry of the system. 
By comparing to simulations with smaller or larger edge potentials, we verify that their  presence does not influence the magnitude of the order parameter.

\begin{figure}[tb]
\includegraphics[width=1\linewidth]{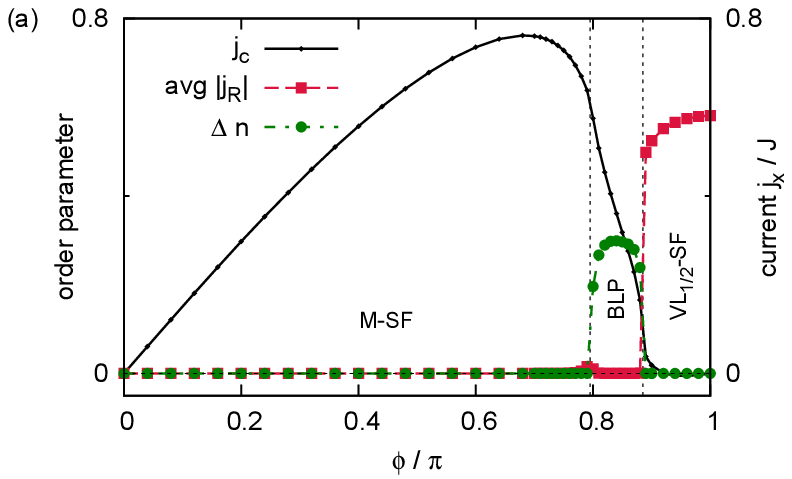}
\includegraphics[width=1\linewidth]{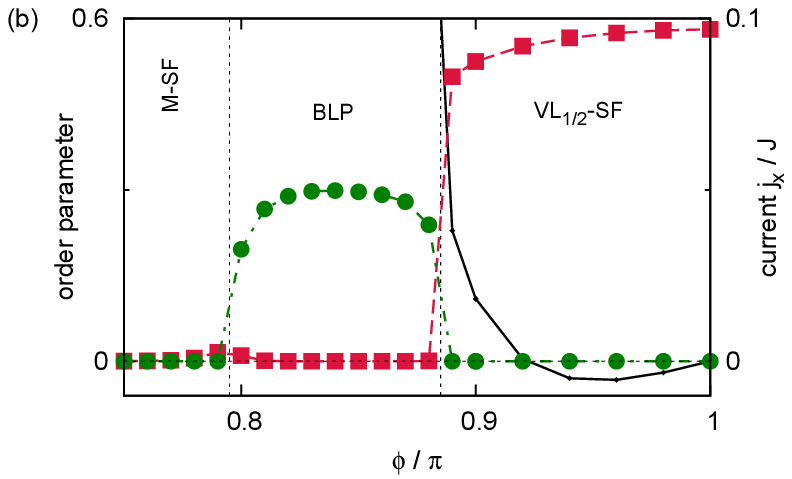}
\caption{Biased-ladder phase: (a) Order parameter $\Delta n$ (imbalance) for the BLP phase, the chiral current $j_c$, and the averaged rung-current ${\rm avg} |j_R|$
for $\rho=0.8$, $U/J=2$ and $J_\perp/J=3$ (DMRG calculation, $L=120$ rungs). (b) Enlarged view  of the BLP-to-VL$_{1/2}$ transition region. Again, 
in the VL$_{1/2}$-SF phase, the chiral current reverses its sign.}
	\label{fig:blp_cut}
\end{figure}

We identify the phase transition by a sharply increasing particle density imbalance between the legs $\Delta n$, as presented in Fig.~\ref{fig:blp_cut}. The data
 are consistent with a second-order Ising-type transition between the M-SF and the BLP phase, as one might expect from  the spontaneous  breaking
of a $Z_2$ symmetry.
Still, a weak first-order nature of the transition cannot be excluded.

\section{Summary}
\label{sec:sum}

In summary, we presented an extensive study of the ground-state physics of repulsively interacting bosons on a two-leg ladder in the presence of a 
uniform Abelian gauge field. In particular, we focused on the discussion of quantum phases with (spontaneously) broken discrete symmetries, 
including various vortex-lattice phases, a charge-density-wave phase at quarter filling and the  biased-ladder phase.
We analyzed the vortex-lattice phases at vortex densities $\rho_v=1/2$, $1/3$ and $1/4$ in detail and studied 
different properties and observables such as the central charge, the structure of local currents and the momentum distribution function in different gauges. 
The vortex density $\rho_v$ can be extracted numerically in several ways, such as from analyzing the local current structure or the momentum distribution. We  furthermore characterized  the various phase transitions between these phases and neighboring ones. We investigated the stability of vortex lattices
 against  including nearest-neighbor interactions on the rungs,  relevant for synthetic-lattice dimension experiments~\cite{Mancini2015, Stuhl2015}.

As we  showed in our previous work \cite{Greschner2015}, vortex-lattice phases may feature  an exotic chiral-current reversal effect. 
We here discussed  how this phenomenon may be understood intuitively for a simplified model of noninteracting bosons with an explicitly broken translational symmetry.
Thus, the effect is clearly related to the effective flux seen by the particles, which either results from 
spontaneously enlarging the unit cell or by constructing models with intrinsically larger unit cells.

From the limit of strong rung-couplings $J_\perp/J$, we may understand the emergence of the CDW phase at quarter filling introduced in Ref.~\cite{Piraud2015}. 
Here, we presented results for its stability as a function of $J_\perp/J$, $\phi$ and also $U/J$. Remarkably, at large values of $U/J$,
 we observe a direct transition to a fully gapped VL$_{1/2}$-MI phase. For strong nearest-neighbor interactions on the rungs (as realizable with a 
synthetic lattice dimension), additional CDW phases are  stabilized. 

Finally, we discussed the properties of the BLP phase starting from an analytical analysis that is set up for  the limit of a dilute Bose gas (i.e., a low filling 
$\rho\to 0$), which allows for an intuitive understanding of the nature of the BLP phase.

Open questions naturally arising from the discussion of our model~\eqref{eq:ham_rung} concern  
its connection to the physics studied extensively in two dimensions such as  the fractional quantum-Hall effect or more specifically, fractional Chern insulators \cite{Wang2012,Liu2012,Sterdyniak2013}. 
In particular, the proposal that Laughlin-like phases exist in this model as conjectured in Ref.~\cite{Grusdt2014,Cornfeld2015,Petrescu2015,Petrescu-aps} offer 
exiting further possibilities for studies of this apparently simple, yet ultimately very rich two-leg ladder model.

\begin{acknowledgments}
We are grateful to N. Cooper, T. Giamarchi, E. Jeckelmann,  A. L\"auchli, M. Lein, G. Roux,  and L. Santos for useful discussions.
S.G. acknowledges support of the Research Training Group (RTG) 1729 and project no. SA 1031/10-1 of the German Research Foundation (DFG).
M.P. was supported by the European Union through the FP7/Marie-Curie grant No. 624033 ("ToPOL") and the FP7/Marie-Curie grant No. 321918 ("FDIAGMC").
I.MC. acknowledges funding from the Australian
Research Council Centre of Excellence for Engineered Quantum Systems and grant number CE110001013.
F.H.-M. and U.S. acknowledge support from the DFG (Research Unit FOR 2414) via grants no. HE 5242/4-1 and SCHO 621/11-1.
TV was supported in part by the National Science Foundation under the Grants NSF DMR-1206648.
Simulations were carried out on the cluster system at the Leibniz University of Hannover, Germany, and the Arnold Sommerfeld Center for Theoretical Physics
at LMU Munich, Germany.
This research was supported in part by the National Science Foundation under Grant No. NSF PHY11-25915. 
\end{acknowledgments}

\section{Appendix}

\begin{figure}[tb]
\includegraphics[width=0.49\columnwidth]{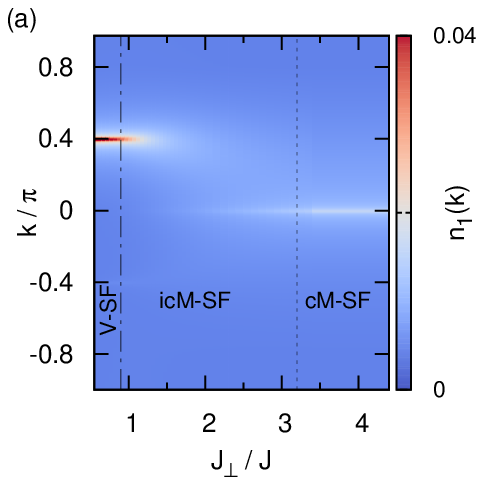}
\includegraphics[width=0.49\columnwidth]{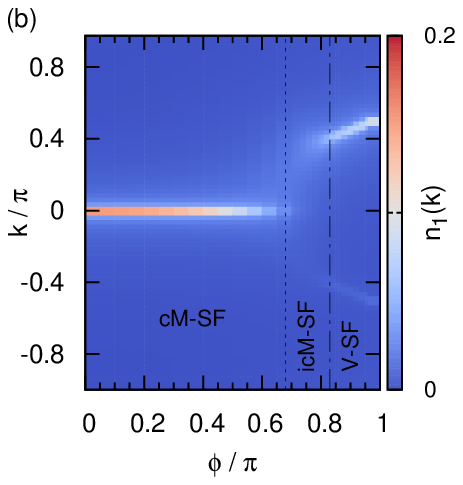}
\caption{Momentum distribution $n_1(k)$ for $U/J\to\infty$, $\rho=0.2$ as a function of (a) $J_\perp/J$ for $\phi/\pi=0.8$ and (b) the flux $\phi/\pi$ for $J_\perp=1.6 J$.
The dotted line indicates the value of $J_\perp$(flux) beyond which we observe an incommensurate behavior (icM-SF) inside the Meissner phase. The transition into the 
vortex-liquid state (V-SF) occurs at $(J_{\perp}^c,\phi_c)$, indicated by the dot-dashed line.
}
	\label{fig:Ap1}
\end{figure}

\begin{figure}[tb]
\includegraphics[width=0.49\columnwidth]{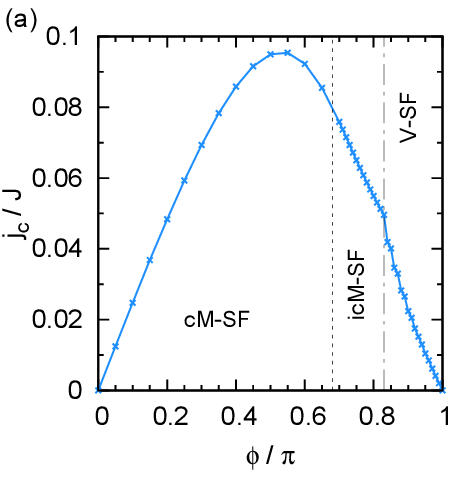}
\includegraphics[width=0.49\columnwidth]{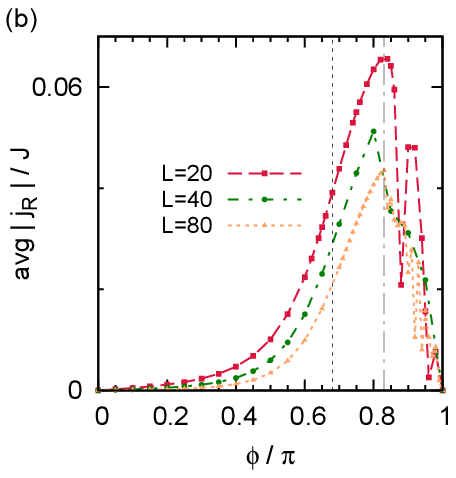}
\caption{(a) Chiral current $j_c$  and (b) average rung currents ${\rm avg}| j_R |$ for $U/J\to\infty$, $\rho=0.2$ and $L=80$ as a function of the flux $\phi/\pi$ for $J_\perp=1.6 J$.
For small values of $\phi<\phi_c$ (dot-dashed line), we are in the Meissner phase. For $ \phi_{\rm ic} < \phi < \phi_c$, we observe two broad maxima in the momentum
distribution $n_1(k)$ shown in Fig.~\ref{fig:Ap1}(b).
}
	\label{fig:Ap2}
\end{figure}

\begin{figure}[tb]
\includegraphics[width=0.99\columnwidth]{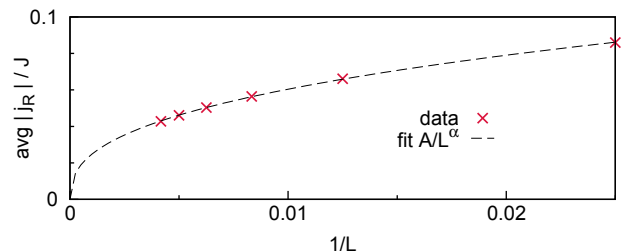}
\caption{Scaling of the average rung current ${\rm avg} |j_R| /J$ with the system size $L$ for $\phi=0.75\pi$, $\rho=0.2$ and $J_\perp=1.6J$. A fit to the data yields $\alpha =0.39$} 
	\label{fig:Ap2c}
\end{figure}

\begin{figure*}[tb]
\includegraphics[width=0.55\columnwidth]{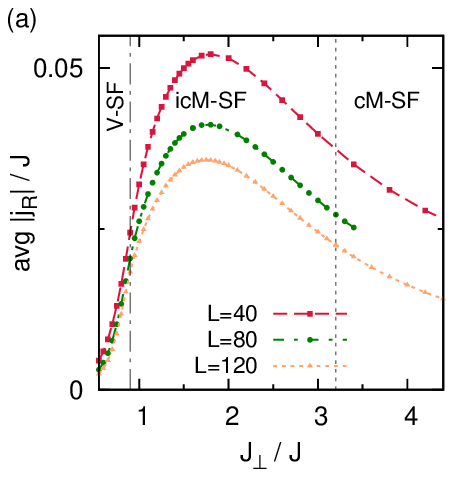}
\includegraphics[width=0.55\columnwidth]{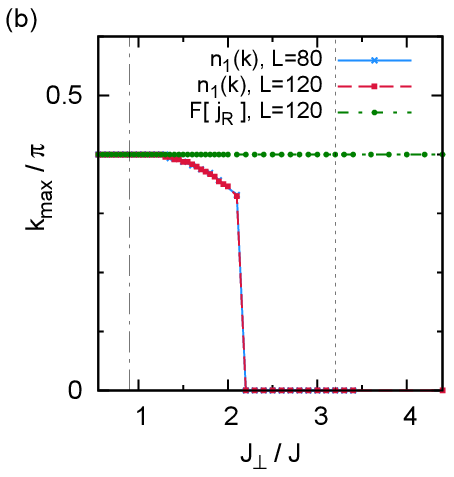}
\includegraphics[width=0.55\columnwidth]{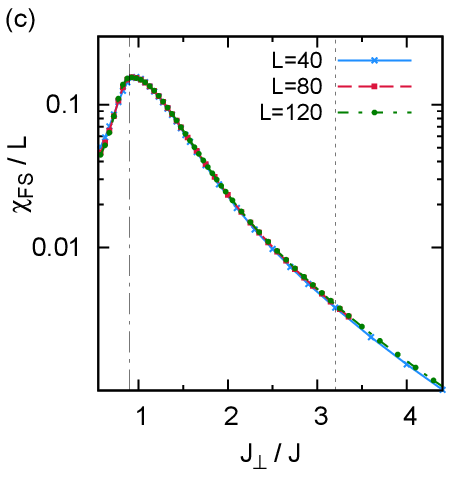}
\includegraphics[width=0.55\columnwidth]{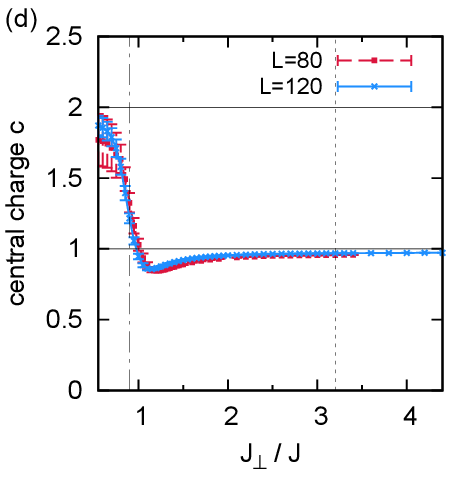}
\includegraphics[width=0.55\columnwidth]{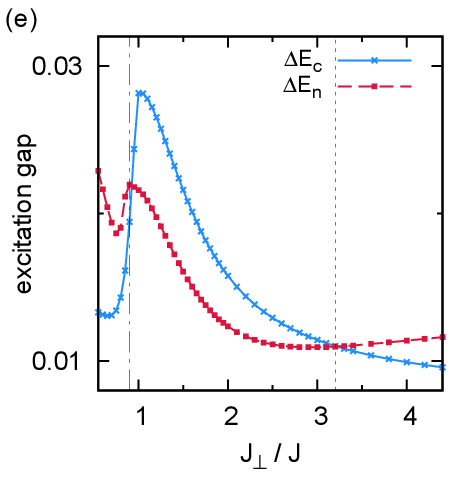}
\includegraphics[width=0.55\columnwidth]{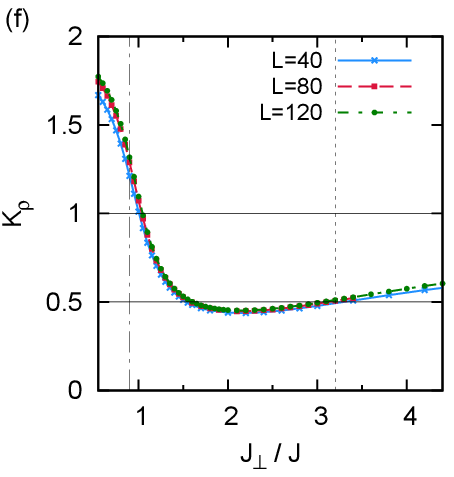}
\caption{(a) Average rung currents ${\rm avg}| j_R |$,
(b) position $k_{\rm max}$ of the maximum of $n_1(k)$ and (c) fidelity susceptibility $F[j_R](k)$,
(d) central charge $c$, (e) excitation gaps for $L=80$, and 
(f) Luttinger-liquid parameter $K_\rho$, as a function of $J_\perp/J$
for $U/J\to\infty$, $\rho=0.2$ and  $\phi/\pi=0.8$.
For small values of $J_\perp <J_{\perp}^c$ (dot-dashed line), we are in the Meissner phase. Upon increasing $J_\perp$, we observe two broad maxima in the momentum
distribution $n_1(k)$ shown in Fig.~\ref{fig:Ap1}(a). The transition into the vortex-liquid phase takes place at $J_\perp^c/J  \sim 3.2 $ (dot-dashed line).
}
\label{fig:Ap3}
\end{figure*}

In this appendix, we address the problem of defining the vortex density in certain parameter regimes. 
For the sake of simplicity, we focus on the limit of hard-core bosons. 
We consider the case of a low particle filling $\rho=0.2$ and increase the flux starting from the Meissner phase (or alternatively, we 
increase $J_{\bot}$ starting from the vortex-liquid state). 

We start by analyzing the momentum distribution function in the symmetric leg gauge of Eq.~\eqref{eq:ham_leg}, as a function of either $J_\perp/J$ (Fig.~\ref{fig:Ap1}(a)) or $\phi$ (Fig.~\ref{fig:Ap1}(b)).
The transition to the vortex-liquid state V-SF is indicated by the dashed-dotted line.
From Fig.~\ref{fig:Ap1}, we see that below a certain value of $J_{\perp}^{\rm ic}$ or  flux $\phi_{\rm ic}$ indicated by the dotted lines, the momentum distribution function becomes blurred in the Meissner phase.  Anticipating the results of the following discussion we denote the two regions as commensurate (cM-SF) and incommensurate (icM-SF) regions of the 
Meissner-superfluid phase.
Thus, $n_1(k)$ ceases to be
sharply  peaked at zero momentum, which we otherwise would expect   for the Meissner phase if we linked the position of the maximum  
of the momentum distribution to the vortex density.
 In addition to the shallow maximum at $k=0$, another maximum at a $k_m \neq 0$ appears in the momentum distribution function for 
$ J_{\perp}^{\rm ic}< J_{\perp}  <J_{\perp}^c $  and 
$ \phi_{\rm ic}< \phi <\phi_c $  at a $k=k_m\neq 0$, respectively, 
and the weight of the momentum distribution continuously shifts from $k=0$ to $k=k_m$ when moving towards the vortex-liquid phase. 
In the same parameter regime, we observe strong modulations of local particle densities and currents, which extend deep into the Meissner phase 
and diminish smoothly when departing from the phase transition $(J_\perp^{c}, \phi_c)$ (dot-dashed lines in the figure) from the vortex-liquid states into  the Meissner regime. 
These oscillations are the combined effect of open boundaries and finite system sizes and die out with increasing the system size as can be seen in 
Figs.~\ref{fig:Ap2}(b), ~\ref{fig:Ap2c} and ~\ref{fig:Ap3}(a). As shown in Fig.~\ref{fig:Ap2c} the numerical data indicates that the average rung-current ${\rm avg} |j_R| /J$ vanishes as $\sim 1/L^\alpha$ as expected for boundary driven effects. The vortex density extracted from the Fourier transform of the rung currents shows a plateau at $k_{\rm max}=0.4$ as depicted in 
Fig.~\ref{fig:Ap3}(b). 
That plateau extends deep into the Meissner phase and its presence can thus not be used as an unambiguous measure of vortex density. 

Most importantly, we could not find any trace of an actual phase transition between the icM-SF regime with multiple peaks in the momentum distribution 
and the conventional cM-SF phase with its single maximum in the momentum distribution at $k=0$. In particular, the fidelity susceptibility (see Fig.~\ref{fig:Ap3}(c)) is 
featureless and the block entanglement entropy does not indicate the presence of a conventional second-order phase transition (see Fig.~\ref{fig:Ap3}(d)). 
The blurring of the momentum distribution may indicate that single-particle excitations become either gapped, which would result in a phase that is thermodynamically distinct
 from the 
Meissner phase, or that the single-particle correlations still decay algebraically, but much slower than in the Meissner phase at either small $J_\perp/J$ or $\phi$  (see Fig.~\ref{fig:Ap3}(f)). 
 Even if we did not observe any drastic change in the ground-state characteristics between the Meissner phase and the regime with pronounced finite-size modulations realized 
close to the boundary of the vortex-liquid state, we observe a distinct level crossing in excited states with negligible finite-size effects as shown in  Fig.~\ref{fig:Ap3}(e). 
In the bulk of the Meissner phase, the lowest excitation is a single-particle excitation
$\Delta E_c = \left(E_0(L,N-1) - 2 E_0(L,N) + E_0(L,N+1)\right)/2$
with $E_m(L,N)$ being the $m$-th eigenenergy of a system of $L$ rungs and $N$ particles, while near the boundary between the Meissner and vortex-liquid phases,
 the lowest excitation is a charge-neutral excitation from the subspace that has  the same number of particles as the ground state,
$\Delta E_n(L,N) = E_1(L,N) - E_0(L,N)$. We mark this level-crossing position, which coincides with $K_\rho=0.5$ (compare Figs.~\ref{fig:Ap3}(e) and (f)), with a dotted line in Figs.~\ref{fig:Ap1}-\ref{fig:Ap3}.

Based on our current data, it remains  unclear whether we are dealing with a thermodynamically distinct state from the Meissner state such as, e.g., a $\nu=1/2$ 
Laughlin state \cite{Petrescu2015,Petrescu-aps}, where the vortex density is pinned to two times the particle density.


\bibliography{references}

\end{document}